\documentclass[10pt,english,prd,superscriptaddress,nofootinbib,preprintnumbers,showpacs,showkeys,floatfix]{revtex4}

\usepackage[latin1]{inputenc}
\usepackage[T1]{fontenc}
\usepackage{lmodern}
\usepackage{amsmath}
\usepackage{amssymb}
\usepackage{graphicx}
\usepackage{caption}
\usepackage{subcaption}
\usepackage{float}
\usepackage{esint}
\usepackage{bm}
\usepackage{amsmath}

\usepackage{dcolumn}
\usepackage{babel}
\usepackage{csquotes}
\usepackage{color}
\usepackage{slashed}
\usepackage{simplewick}
\usepackage{amsmath,latexsym}
\usepackage{slashed}
\usepackage{doi}
\usepackage{hyperref}
\hypersetup{
    colorlinks,
    breaklinks,
    citecolor=[rgb]{0,0.0,1.0},
    urlcolor=[rgb]{0.0,0.0,1.0},
    linkcolor=[rgb]{0,0.5,0.9}
}

\begin{document}

\title{The End of the Road for Bulk Fields in Warped Randall-Sundrum Braneworlds}
\author{G. Alencar}
\email{geova@fisica.ufc.br}
\affiliation{Departamento de F\'isica, Universidade Federal do Cear\'a, Caixa Postal 6030, Campus do Pici, 60455-760 Fortaleza, Cear\'a, Brazil.}
\author{R. S. Almeida}
\email{rodrigoalmeida@fisica.ufc.br}
\affiliation{Departamento de F\'isica, Universidade Federal do Cear\'a, Caixa Postal 6030, Campus do Pici, 60455-760 Fortaleza, Cear\'a, Brazil.}
\author{R. N. Costa Filho}
\email{rai@fisica.ufc.br}
\affiliation{Departamento de F\'isica, Universidade Federal do Cear\'a, Caixa Postal 6030, Campus do Pici, 60455-760 Fortaleza, Cear\'a, Brazil.}
\author{T. M. Crispim}
\email{tiago.crispim@fisica.ufc.br}
\affiliation{Departamento de F\'isica, Universidade Federal do Cear\'a, Caixa Postal 6030, Campus do Pici, 60455-760 Fortaleza, Cear\'a, Brazil.}

\author{Francisco S. N. Lobo} 
\email{fslobo@ciencias.ulisboa.pt}
\affiliation{Instituto de Astrof\'{i}sica e Ci\^{e}ncias do Espa\c{c}o, Faculdade de Ci\^{e}ncias da Universidade de Lisboa, Edif\'{i}cio C8, Campo Grande, P-1749-016 Lisbon, Portugal}
\affiliation{Departamento de F\'{i}sica, Faculdade de Ci\^{e}ncias da Universidade de Lisboa, Edif\'{i}cio C8, Campo Grande, P-1749-016 Lisbon, Portugal}

\date{\today}

\begin{abstract}
In this manuscript we generalize Ref. \cite{Alencar:2024lrl} and derive a complete set of local consistency conditions for bulk fields in braneworld scenarios with an arbitrary number of dimensions. This provides the first fully local and dimension-independent generalization of all known criteria for bulk fields. Within this framework, we show that a free scalar field is consistent and localized, whereas minimally and non-minimally coupled Maxwell fields violate the conditions, leading to a no-go theorem valid in any dimension. For nonlinear electrodynamics, we find that only the model $L(F)=b\sqrt{F}$ admits a consistent and normalizable zero mode, and that among $p-$forms, consistency occurs solely for the free $0-$form. We also demonstrate that Dirac fermions, with or without Yukawa terms, are inconsistent within this framework and therefore cannot propagate in the bulk. Our local approach makes explicit that these conclusions do not depend on any particular internal geometry or warp factor: previously known results arise merely as special cases of a broader and strictly local structure, highlighting the universality of the constraints derived here.
\end{abstract}


\maketitle

\begingroup
\small
\renewcommand{\baselinestretch}{0.87}\normalsize
\tableofcontents
\endgroup

\section{Introduction}

Extra-dimensional constructions based on branes now occupy a central place in high-energy theory, largely because they provide novel perspectives on fundamental problems such as the hierarchy problem and the unification of forces \cite{Gogberashvili:1998vx, Gogberashvili:1999tb, Randall:1999ee, Randall:1999vf}. In this approach, the observable universe is identified with a four-dimensional hypersurface residing within a higher-dimensional bulk, and the presence of extra spatial directions plays a crucial role in shaping gravitational phenomena across widely separated length scales.
Configurations in which our universe is realized as a brane embedded in a higher-dimensional spacetime with a warped, non-factorizable geometry have attracted particular attention, as they offer fresh avenues to tackle several unresolved issues in contemporary physics (see, for example, the reviews \cite{Maartens:2010ar,Raychaudhuri:2016kth}). These constructions are commonly referred to as Randall--Sundrum (RS) scenarios and were initially formulated in five dimensions, involving either a single brane or a pair of branes. In the version containing two branes, often labeled the RS type I model, Randall and Sundrum demonstrated how the hierarchy problem could be addressed through the geometric properties of the extra dimension.

Within the context of RS constructions, there exist two widely adopted strategies for introducing matter degrees of freedom. One of these takes the viewpoint that all matter sources are entirely localized on the brane, so that the corresponding energy--momentum tensor has no support in the bulk. Under this assumption, the gravitational dynamics perceived on the brane are derived by projecting the higher-dimensional Einstein equations onto the brane via the Gauss--Codazzi relations. This procedure, formulated by Shiromizu, Maeda, and Sasaki \cite{Shiromizu_2000}, leads to an effective four-dimensional set of gravitational equations in which bulk effects manifest themselves through extra contributions. Among these is the term associated with the projected Weyl tensor, which encapsulates the influence of the higher-dimensional gravitational field on brane physics. Although this framework offers a systematic way to extract a four-dimensional gravitational theory from the higher-dimensional setup, it does not address the underlying mechanism that would dynamically localize matter fields on the brane.

An alternative and more microscopic viewpoint assumes that matter degrees of freedom are not restricted a priori to the brane, but instead propagate throughout the higher-dimensional bulk. In this setting, one begins by formulating the field equations in the full spacetime and subsequently investigates whether the zero-modes of fields admit a dynamical confinement onto the brane. The usual requirement for such localization is that, once the equations of motion are solved for the zero mode of a given field, the resulting action remains finite when integrated over the extra dimensions.
This line of reasoning is particularly important for clarifying the zero-mode localization mechanisms of fundamental fields, such as gauge fields and fermions, and it underpins the construction of viable higher-dimensional scenarios. Nonetheless, when gravitational effects are taken into account, exact confinement of zero modes is not guaranteed and may even be obstructed \cite{Fichet:2019owx}. Moreover, even in situations where zero-mode localization is achieved, a finite extra-dimensional integral by itself does not suffice to guarantee overall consistency once the coupled dynamics of gravity and matter fields are fully considered.

In the single-brane realization, often referred to as the RS type II scenario, the usual notion of compactifying the extra dimension is replaced by a setup in which the additional dimension extends infinitely. The spacetime geometry is described by a warped line element of the form
\begin{equation}
	ds^2 = e^{2A(y)} g_{\mu\nu}(x)\, dx^{\mu} dx^{\nu} + dy^2,
\end{equation}
where $g_{\mu\nu}(x)$ denotes the induced metric on the brane and $y$ labels the extra spatial direction. This warped, non-factorizable structure leads to several noteworthy consequences.
To illustrate this, consider the five-dimensional Einstein--Hilbert action
\begin{equation}
	\mathcal{S}_{EH} = \int d^5x \sqrt{_{(5)}g(x,y)}\, R_{(5)}(x,y),
\end{equation}
with $R_{(5)}$ the five-dimensional Ricci scalar. Hereafter, the labels ``$(5)$'' and ``$(4)$'' will be used to distinguish five- and four-dimensional geometric quantities.

For the metric above, both the determinant of the metric and the Ricci scalar decompose according to
\begin{equation}
	\sqrt{_{(5)}g(x,y)} = e^{4A(y)} \sqrt{_{(4)}g(x)}, 
	\qquad 
	R_{(5)}(x,y) = e^{-2A(y)} R_{(4)}(x) - 20 A'(y)^2 - 20 A''(y),
\end{equation}
where primes denote derivatives with respect to the extra coordinate $y$. Substituting these expressions into the action yields
\begin{equation}
	\mathcal{S}_{EH} = \int dy\, e^{2A(y)} \int d^4x \sqrt{_{(4)}g(x)}\, R_{(4)}(x) + \ldots \; .
\end{equation}
The key observation is that the action separates into a product of an integral over the extra dimension and a four-dimensional Einstein--Hilbert term. Consequently, provided that the integral over $y$ converges, one obtains a well-defined effective four-dimensional theory of gravity. Upon including the brane contributions as sources and solving the resulting equations of motion, the warp factor $A(y)$ is determined explicitly, and one finds that the integral over the infinite extra dimension is indeed finite.

Once the gravitational background has been specified, it is natural to investigate the behavior of additional bulk fields. Consider a generic five-dimensional field $\Phi(x,y)$ acted upon by some differential operator $\hat{O}$ defined in the full spacetime. A standard step is to assume a separable form for the field, $\Phi(x,y)=\psi(y)\,\phi(x)$, and to decompose the operator accordingly as $\hat{O}=\hat{O}_{4d}+\hat{O}_{y}$. This leads to the relation
\begin{equation}
\frac{\hat{O}_{4d}\phi(x)}{\phi(x)} = - \frac{\hat{O}_{y}\psi(y)}{\psi(y)} ,
\end{equation}
which implies the pair of eigenvalue equations
\begin{equation}
\hat{O}_{4d}\phi(x) = -m^2 \phi(x), 
\qquad 
\hat{O}_{y}\psi(y) = m^2 \psi(y).
\end{equation}
The spectrum of solutions to the equation along the extra dimension therefore determines the mass eigenvalues observed on the brane.
Applying this separation to the action of the bulk field yields
\begin{equation}
	\mathcal{S}_{\Phi} = \int d^5x \sqrt{-g}\,\mathcal{L}_5(x,y)
	= \int dy\, f\!\left(\psi, e^{A}\right)\int d^4x \sqrt{_{(4)}g(x)}\, \mathcal{L}_4(x),
\end{equation}
where the function $f(\psi,e^{A})$ depends on the profile of the mode in the extra dimension and on the warp factor. When the integral over $y$ converges, the resulting four-dimensional action is finite and well defined. This mechanism, by which bulk fields give rise to effective four-dimensional degrees of freedom concentrated near the brane, is commonly referred to as field localization.

A particularly important aspect of this analysis concerns the behavior of massless modes, corresponding to $m^2=0$, since their localization is essential for reproducing familiar four-dimensional physics on the brane, such as the presence of massless gauge fields. Early investigations, however, revealed significant obstacles in this direction. In the context of the RS type I setup, it was demonstrated that free gauge fields cannot be consistently localized on the brane \cite{Davoudiasl:1999tf,Pomarol:1999ad}, highlighting a nontrivial tension between extra-dimensional geometry and field dynamics.
Motivated by these issues, Kaloper and collaborators studied a more general configuration consisting of a $d$-dimensional brane embedded in a $(D=d+1)$-dimensional bulk spacetime \cite{Kaloper:2000xa}. Their analysis showed that $p$-form fields satisfying the condition $p<(d-2)/2$ possess normalizable zero modes, meaning that the corresponding integrals over the extra dimension converge and the fields appear effectively localized on the brane. This result provided a broader classification of which types of fields can be trapped by warped geometries.
Nevertheless, from a more complete perspective in which gravitational backreaction and matter dynamics are treated on equal footing, the mere convergence of the extra-dimensional integral does not automatically guarantee a fully consistent effective theory. Additional consistency conditions may arise once the coupled system of gravity and fields is taken into account, indicating that localization based solely on normalizability is a necessary but not always sufficient criterion.

Indeed, the necessity of going beyond simple normalizability arguments was first emphasized by Duff and collaborators in their investigation of $p$-form fields within RS type II scenarios. They argued that a consistent localization mechanism must not only yield a finite effective action after integration over the extra dimension, but must also respect the full set of the Einstein field equations, including the backreaction induced by the localized fields on the spacetime geometry. When this stronger gravitational consistency requirement is imposed on $p$-form fields, their analysis shows that only the scalar ($0$-form) and its Hodge dual satisfy all of the Einstein equations simultaneously \cite{Duff:2000se}. This result effectively excludes all other $p$-forms, even though earlier work had established that their extra-dimensional integrals are finite \cite{Kaloper:2000xa}.
Shortly thereafter, Gibbons and coauthors extended this reasoning to the RS type I setup by explicitly incorporating backreaction effects \cite{Gibbons:2000tf}. They derived a set of nontrivial consistency relations expressed as integrals over the compact extra dimension, which we refer to as the ``Global Sum Rule'' (GSR). One notable implication of these constraints is that the Goldberger--Wise stabilization mechanism \cite{Goldberger:1999uk} fails to satisfy the GSR, despite the fact that scalar fields also lead to convergent integrals. Subsequent studies generalized and refined these ideas, introducing further consistency conditions that impose additional restrictions on bulk field dynamics and on the localization of their zero modes \cite{Gibbons:2000tf, Leblond:2001xr, Freitas:2020mxr}. Collectively, these results clarify which types of fields can be consistently confined to the brane within a fully self-consistent gravitational and field-theoretical framework.

In the years following these developments, a wide variety of mechanisms have been proposed to address the localization of both fermionic and bosonic fields in braneworld scenarios \cite{Bajc:1999mh, Landim:2015paa, Liu:2008pi, Koley01, Zhao:2010mk, Li:2010dy, Liu:2009ve, Guo, Liu:2009dw, Brihaye:2012um, Oda:2000kh, Ringeval:2001cq, Mendes, Melfo, Castro, Barbosa-Cendejas:2015qaa, Kehagias, Zhao:2009ja, Farokhtabar:2016fhm, Almeida:2009jc, Bazeia:2004dh, Guerrero:2009ac, Fu:2011pu, Barut, Oda02, Liu01, Liu03, Liu04, Liu02, Oda03, Oda06, Midodashvili, Oda05, Batell, Casadio, Alencar01, Germani, GLandim01, GLandim02, Oda07, Ghoroku, Alencar02, Zhao, Fu01, Chumbes, Alencar03, Cruz01, Cruz02, Flachi, Freitas, Costa01, Costa02, Zhao02, Landim01, Landim02, Landim03, Sui, Torrealba, Giovannini, Fu02, GAlencar}. Most of these constructions rely on introducing nontrivial couplings between matter fields and background scalar fields, typically accompanied by adjustable parameters that control localization properties. From a theoretical standpoint, however, the presence of arbitrary free parameters is generally viewed as a drawback.
In an effort to reduce this arbitrariness, the gravitational consistency condition originally proposed by Duff \emph{et al.} was extended in Refs.~\cite{Freitas:2020mxr, Freitas:2020vcf, Freitas:2022whn}. When applied to the existing localization models, these generalized constraints were shown to either fix the previously free parameters or to exclude entire classes of constructions altogether. Nonetheless, those analyses did not fully incorporate the implications of treating gravitational dynamics and matter fields as a coupled system.
A complete set of constraints was obtained more recently by some of the present authors in Ref.~\cite{Alencar:2024lrl}. That work establishes a collection of no-go conjectures that impose stringent restrictions on the localization of zero modes for bulk fields in five-dimensional RS braneworlds, with particularly strong consequences for gauge and spinor fields. Unlike conventional approaches, the analysis does not depend on specifying detailed equations of motion, which makes the results applicable to a broad class of braneworld models. These no-go conditions uncover fundamental obstructions to field localization, casting doubt on the viability of confining certain fields to the brane. In particular, it is shown that known five-dimensional models do not achieve consistent localization of gauge and fermionic fields, and that the resulting effective brane Lagrangian cannot be conformally invariant.

Thus, the motivations of the present work stem from the fact that the analysis of higher-dimensional braneworlds and more general gravitational backgrounds remains largely unexplored. Building on this, the aim of the paper is to extend the results of Ref.~\cite{Alencar:2024lrl} to arbitrary spacetime dimensions and bulk fields. Specifically, we derive a fully local and dimension-independent set of consistency conditions for bulk fields, providing a comprehensive generalization of all known criteria. This framework allows us to identify which fields can be consistently localized on the brane, revealing, for instance, that free scalar fields are admissible while minimally and non-minimally coupled Maxwell fields, most $p$-forms, and Dirac fermions are prohibited in any dimension. The local nature of our approach ensures that these conclusions hold independently of the specific warp factor or internal geometry, highlighting the universality of the derived constraints.

The paper is organized as follows: In Sec.~\ref{SecII}, we provide a detailed review of the main consistency conditions that arise in braneworld constructions. This section opens with a discussion of the Global Sum Rules, followed by a presentation of the consistency condition introduced by Duff \emph{et al.} We then review its subsequent generalizations, introduce the local sum rules proposed by Alencar \emph{et al.}, and conclude with a formulation of the Gibbons \emph{et al.} consistency condition extended to arbitrary spacetime dimensions. 
In Sec.~\ref{SecIII}, we derive local sum rules for bulk fields in general dimensions. We first describe the background solutions and then analyze the resulting consequences for the behavior and localization of bulk fields. Section~\ref{SecIV} is devoted to explicit applications of the formalism to different types of matter fields, including scalar fields, bosonic fields, and fermionic fields. In the fermionic case, we separately examine spinor fields in odd-dimensional spacetimes of the form $D=2k+3$ and in even-dimensional spacetimes of the form $D=2k+2$. Finally, Sec.~\ref{SecV} contains a summary of the main results along with our concluding remarks.

\section{Review of braneworld consistency conditions}\label{SecII}

Within the framework of Randall--Sundrum (RS) constructions, the confinement of zero modes for gravitational and scalar degrees of freedom on a positive-tension brane is well established \cite{Randall:1999vf,Bajc:1999mh}. In contrast, attempts to localize massless vector fields encounter significant obstacles due to the conformal invariance inherent to the simplest vector field action, which prevents the standard localization procedures from producing a finite effective action. This difficulty has been analyzed in detail and identified as a fundamental obstruction in Ref.~\cite{Davoudiasl:1999tf}. In this work, we present an approach that overcomes both the challenges associated with vector field localization and the issues linked to constructing a phenomenologically viable braneworld scenario.

\subsection{Global Sum Rules}

Some years ago, Gibbons and collaborators \cite{Gibbons:2000tf} demonstrated that all warped braneworld constructions must obey a set of specific consistency requirements, given by
\begin{equation}\label{sumrule}
\left(A'e^{nA}\right)'=\frac{2\pi G_{5}}{3}e^{nA}\left(T_{\mu}{}^{\mu}+(2n-4)T_{5}{}^{5}\right)-\frac{1-n}{12}e^{(n-2)A}R_{g}.
\end{equation}
The authors integrate the above equation in both sides to obtain the sum rules
\begin{equation}\label{GSR}
\oint e^{nA}\left(T_{\mu}{}^{\mu}+(2n-4)T_{5}{}^{5}\right)=\frac{1-n}{8\pi G_{5}}\oint e^{(n-2)A}R_{g}.
\end{equation}

Next, they consider the example of a scalar field
\begin{equation}\label{scalarfield}
S_b={1\over 2}\int d^4 x\int_{-\pi}^\pi dy\, \sqrt{-g}\,
\left(g^{AB}\partial_A \Phi \partial_B \Phi - m^2 \Phi^2\right),
\end{equation}
with energy-momentum tensor
\begin{equation}\label{scalarfield2}
    T_{MN}  =\partial_{M}\phi\partial_{N}\phi-g_{MN}(\frac{1}{2}g^{PQ}\partial_{P}\phi\partial_{Q}\phi+V(\Phi)),
\end{equation}
which implies that 
 \begin{equation}
 \oint e^{n
A}
  \left( (4-n) \Phi'\cdot \Phi'   + 2n V(\Phi) +4
   \sum_{\alpha} \lambda_\alpha(\Phi) \delta(y-y_\alpha)\right)=4 M^3(1-n) \oint e^{(n-2)A} R_g.
\end{equation}

The authors then evaluate the expression for specific values of $n$ while setting $R_g=0$. In particular, for $n=0$, $n=1$, and $n=4$, the results are, respectively,
\begin{align}
 \label{n=0}
 & \sum_{\alpha} \lambda_\alpha(\Phi_\alpha)  + \oint
\Phi'\cdot \Phi'    =0 \,,
\\
 \label{n=1}
&4\sum_{\alpha}e^{ A} \lambda_\alpha(\Phi_\alpha) +\oint  e^{ A}
  \left(  3  \Phi'\cdot \Phi'   + 2 V(\Phi) \right) =0 \,,
  \\
&\sum_{\alpha} e^{ 4 A}\lambda_\alpha(\Phi_\alpha)+ 2\oint e^{4 A}
   V(\Phi)   = 0 \,.
 \label{n=4}
\end{align}
Under the original Randall--Sundrum fine-tuning condition, $\sum_{\alpha} \lambda_\alpha(\Phi_\alpha)=0$, Eq.~(\ref{n=0}) then implies that $\Phi$ must be constant. As a consequence, smooth type I RS models, such as the KMPRS construction, are rendered inconsistent \cite{Kogan:1999wc}. More explicitly, assuming $\Phi$ is constant and taking the potential $V$ to be equal to a cosmological constant, $V=\Lambda$, Eqs.~(\ref{n=1}) and (\ref{n=4}) lead directly to the relation $\Lambda = k \lambda$.

The authors further examine mechanisms for stabilizing the extra dimension. In the case of the Goldberger--Wise (GW) approach \cite{Goldberger:1999uk}, these introduce, alongside Eq.~(\ref{scalarfield}), an additional interaction term of the form 
\begin{equation} \label{GWaction}
S_h = -\int d^4 x \sqrt{-g_h}\, \gamma_h \left(\Phi^2 - v_h^2\right)^2,
\qquad
S_v = -\int d^4 x \sqrt{-g_v}\, \gamma_v \left(\Phi^2 - v_v^2\right)^2.
\end{equation}
With this, the only change to Eq.~(\ref{n=0}) is 
\begin{eqnarray}
\int _{-\pi}^{\pi}dy\; \Phi'\cdot \Phi' \, + \, \left[ \lambda_{h} +
\lambda_{v}
+\gamma_h \left(\Phi^2(0) - v_h^2\right)^2+\gamma_v \left(\Phi^2(\pi) -
v_v^2\right)^2\right] = 0 \,. \label{GW}
\end{eqnarray}
Given that $\lambda_{h} + \lambda_{v} = 0$, the constraint discussed above cannot be fulfilled, rendering the GW stabilization mechanism inconsistent. The authors then turn to the alternative approach of DeWolfe, Freedman, Gubser, and Karch \cite{DeWolfe:1999cp}. To address the issue, they employ the ``superpotential'' method, defining the scalar profile through $\Phi' = \frac{1}{2} \frac{\partial W}{\partial \Phi}$ with the superpotential chosen as $W(\Phi) = \frac{3}{L} - b \Phi^2$. Within this framework, Eq.~(\ref{n=0}) can be satisfied because
\begin{equation}
    \lambda_1(\Phi_1)+\lambda_2(\Phi_2)=- \oint
\Phi'\cdot \Phi'=- \oint W' =b(\Phi_2^2-\Phi_1^2) \,.
\end{equation}

We note that the analysis in Ref.~\cite{Gibbons:2000tf} applies the above conditions in an averaged sense, by performing an integration over the extra dimension. This procedure is commonly referred to as the ``Global Sum Rule'' (GSR). An additional observation is that the GSR was not extended to include other bulk fields in their study.

\subsection{Duff \emph{et al.} consistency condition}

A new type of consistency requirement for $p$-form fields was introduced by Duff and collaborators \cite{Duff:2000se}. In their setup, one considers a $d$-dimensional brane embedded in a $(D=d+1)$-dimensional braneworld, along with a $p$-form field described by the action
\begin{equation}\label{bulkaction}
S_{\rm bulk}=\int d^{d+1}x
\left[-\frac{1}{2(p+1)!}\sqrt{-\hat g}\hat g^{M_{1}N_{1}}
\hat g^{M_{2}N_{2}}\cdots \hat g^{M_{p+1}N_{p+1}}
\hat Y_{M_{1}M_{2}\ldots M_{p+1}}
\hat Y_{N_{1}N_{2}\ldots N_{p+1}}\right].
\end{equation}
where $Y_{M_{1}M_{2}\ldots M_{p+1}}=\partial_{[M_1} X_{M_{2}M_{3}\ldots M_{p+1}]}$.  The authors of Ref. \cite{Kaloper:2000xa} showed that any $p-$form with $p<(d-2)/2$ has a convergent integral and, therefore, are localized over the brane. The procedure can be simplified as follows \cite{Landim:2010pq}. By fixing the gauge, the  equation of motion is given by
\begin{eqnarray}
\partial _{\mu_1 }Y^{\mu_1 \cdots\mu_{p+1}}
+e^{((2p+2-d)A)}\partial _{y}\left[e^{(d-2p)A}{X'}^{\mu_2\cdots\mu_{p+1}}\right]=0 \,.
\end{eqnarray}
We can now separate the $y$ dependence of the field using $X^{\mu_1\cdots\mu_p}\left( x^{\alpha },y\right) =B^{\mu_1\cdots\mu_p}\left(
x^{\alpha }\right) \psi\left( y\right)$  and by defining $Y^{\mu_1\cdots\mu_{p+1}}=\tilde{Y}^{\mu_1\cdots\mu_{p+1} }\psi(y)$, where $\tilde{Y}$ stands for the $d$-dimensional field strength, we get for the zero mode equations
\begin{equation}\label{pformmass}
 \partial _{\mu_1 }\tilde{Y}^{\mu_1 \cdots\mu_{p+1}}=0, \qquad \partial _y\left(e^{- (2p-d){A}} {\psi}'(
y)\right) =0 \,,
\end{equation}
with this, we arrive at the effective action 
\begin{equation}
S_{X}=\int dy \, e^{{(d-2p-2)A}}\psi^{2}\int
d^{d}x\, \left[\tilde{Y}_{\mu_1\cdots\mu_{p+1}}\tilde{Y}^{\mu_1 \cdots\mu_{p+1}}\right]\,,
\end{equation}
and it is easy to see that $\psi=\psi_0$, with $\psi_0$ being a constant, solves the above equation and we have that the integration in the extra dimension is
finite for $p<(d-2)/2$. 

Following this, Duff \emph{et al.} carried out a thorough analysis of the system described above \cite{Duff:2000se}. They noted that the initial result appears to conflict with Hodge duality \cite{Duff:1980qv}, which requires that all dual forms with $p > d/2$ should also yield convergent integrals. To address this issue, they proposed that a bulk $p$-form can induce either a $p$-form or a $(p-1)$-form on the brane. Additionally, they observed that the profile $\psi = e^{(2p-d)A}$ also satisfies Eq.~(\ref{pformmass}). Within this framework, the effective action on the brane takes the form
\begin{equation}
S_{X}=\int dye^{{(2p-d)A}}\psi^{2}\int
d^{d}x[\tilde{Y}_{\mu_1\cdots\mu_{p}}\tilde{Y}^{\mu_1 \cdots\mu_{p}}]\,,
\end{equation}
and the $p>d/2$ also have a convergent integral. 
They then investigated whether the convergence of the extra-dimensional integral alone is sufficient to ensure a consistent system. To this end, the full coupled dynamics of gravity and the $p$-form fields are taken into account. In this context, the energy--momentum tensor is given by
\begin{equation}
    T_{MN}=Y_{MM_{1}M_{2}\ldots M_{p}}Y_{N}^{\;M_{1}M_{2}\ldots M_{p}}-\frac{1}{2(p+1)}g_{MN}Y_{M_{1}M_{2}\ldots M_{p+1}}Y^{M_{1}M_{2}\ldots M_{p+1}}\,.
\end{equation}
They first considered the $p-$form case, and the Einstein equation can be written as
\begin{eqnarray}
\label{eq:kk1eins}
^{(4)}{G}_{\mu\nu}(x)&=&\frac{1}{2\cdot p!}
e^{-p A}\psi^2\Bigl[g^{\mu_{1}\nu_{1}}g^{\mu_{2}\nu_{2}}\cdots g^{\mu_{p}\nu_{p}}
{\tilde{Y}}_{\mu\mu_{1}\mu_{2}\ldots \mu_{p}}
{\tilde{Y}}_{\nu\nu_{1}\nu_{2}\ldots \nu_{p}}\nonumber\\
&&-\frac{1}{2(p+1)}{g}_{\mu\nu}
g^{\mu_{1}\nu_{1}}g^{\mu_{2}\nu_{2}}\cdots g^{\mu_{p+1}\nu_{p+1}}
{\tilde{Y}}_{\mu_{1}\mu_{2}\ldots \mu_{p+1}}
{\tilde{Y}}_{\nu_{1}\nu_{2}\ldots \nu_{p+1}}\Biggr]\,.
\end{eqnarray}
Then, they argued that, since the LHS depends only on $x$, the same must happen to the RHS. With this, and the solution $\psi=\psi_0$ for the $p$-form, they found that only $p=0$ is consistent. For the $(p-1)$-form case, the Einstein equation reduces to  
\begin{eqnarray}
\label{eq:kk1eins2}
^{(4)}{G}_{\mu\nu}(x)&=&\frac{1}{2\cdot (p-1)!}
e^{(d-p-1) A}\psi^2\Bigl[g^{\mu_{1}\nu_{1}}g^{\mu_{2}\nu_{2}}\cdots g^{\mu_{p}\nu_{p-1}}
{\tilde{Y}}_{\mu\mu_{1}\mu_{2}\ldots \mu_{p-1}}
{\tilde{Y}}_{\nu\nu_{1}\nu_{2}\ldots \nu_{p-1}}\nonumber\\
&&-\frac{1}{2p}{g}_{\mu\nu}
g^{\mu_{1}\nu_{1}}g^{\mu_{2}\nu_{2}}\cdots g^{\mu_{p}\nu_{p}}
{\tilde{Y}}_{\mu_{1}\mu_{2}\ldots \mu_{p}}
{\tilde{Y}}_{\nu_{1}\nu_{2}\ldots \nu_{p}}\Biggr] \,,
\end{eqnarray}
and only $p=(d-1)$ is consistent. Therefore, when the coupled
Einstein$-p$-form field equations is imposed, the only $p$-forms that can be bound to the brane are the $0$-form and its dual. This result could not
have been found by substituting $\psi$ into the action and demanding
convergence of the integral, and shows the dangers of substituting it into an action rather than the equations of motion \cite{Duff:1984hn}.

\subsection{Generalization of Duff \emph{et al.} consistency condition}

Some time later, the works of Refs.~\cite{Freitas:2020mxr, Freitas:2020vcf, Freitas:2022whn} extended the consistency condition originally proposed by Duff \emph{et al.} Building on this approach, the authors analyzed the full dynamics of gravity coupled to arbitrary bulk fields by introducing a general energy-momentum tensor, with the goal of reassessing previous conclusions regarding field localization. The central idea relies on the decomposition of the Einstein tensor as
\begin{equation}\label{ETspliting}
	{}^{(5)}G_{\mu\nu}(x) = {}^{(4)}G_{\mu\nu}(x) + \hat{g}_{\mu\nu}(x) \big( 3 A'' + 6 A'^2 \big).
\end{equation}

In the absence of bulk fields, the energy-momentum tensor is given by
\begin{equation}\label{RSvacuum}
	T_{\mu\nu}^{(v)} = - \hat{g}_{\mu\nu} \left( \Lambda + \sum_{\alpha} \lambda_{\alpha} \delta(y - y_\alpha) \right),
\end{equation}
where the superscript $(v)$ denotes the ``vacuum'' contribution. For the standard RS warp factor $A = \kappa |y|$, this leads to the usual fine-tuning relations $\lambda_v = -\lambda_h$ and $\Lambda = -\frac{3 \kappa^2}{8 \pi G_5}$. We now extend this setup to include a general bulk field in addition to the branes and the cosmological constant.

In this extended scenario, the five-dimensional Einstein equations take the form
\begin{equation}\label{EE5}
	{}^{(5)}G_{\mu\nu}(x,y) = 8 \pi G_5 \left( T_{\mu\nu}^{(v)} + T_{\mu\nu}^{(b)}(x,y) \right),
\end{equation}
where $T_{\mu\nu}^{(b)}(x,y)$ represents the energy--momentum tensor of the bulk field, with the superscript $(b)$ indicating ``bulk''. Using the warp factor $A = \kappa |y|$ together with Eqs.~(\ref{ETspliting}), (\ref{RSvacuum}), and (\ref{EE5}), we obtain
\begin{equation}\label{condition1freitas}
	{}^{(4)}G_{\mu\nu}(x) = 8 \pi G_5 \, T_{\mu\nu}^{(b)}(x,y) 
	\quad \Longrightarrow \quad 
	T_{\mu\nu}^{(b)}(x,y) = T_{\mu\nu}^{(b)}(x).
\end{equation}

The relation above must hold because the LHS depends solely on the brane coordinates $x$. This is precisely the approach originally employed by Duff \emph{et al.} for $p$-form fields. The key difference here is that the condition is now extended to include arbitrary bulk fields, not just $p$-forms. In the original analysis by Duff \emph{et al.}, even when the extra-dimensional integrals were finite, all $p$-forms except for the scalar field and its Hodge dual were excluded. We anticipate that this generalized condition will impose analogous restrictions on other types of bulk fields.

Building on this framework, the authors of Refs.~\cite{Freitas:2020mxr, Freitas:2020vcf, Freitas:2022whn} were able to rule out a number of previously proposed models for field localization in type II braneworlds. In particular:
\begin{itemize}
	\item \textbf{Fermions}: In Ref.~\cite{Freitas:2020mxr}, by applying the generalized consistency condition together with the equations of motion, it was shown that the only Yukawa coupling of the form $\lambda f(y) \bar{\Psi} \Psi$ that remains consistent with the full gravitational and matter dynamics is 
	\begin{equation}\label{fermioncoupling}
		f(y) = A'(y).
	\end{equation}
	This condition renders the models of Refs.~\cite{Bajc:1999mh, Oda:2000kh} consistent, while excluding the constructions proposed in Refs.~\cite{Koley01, Ringeval:2001cq, Mendes, Melfo, Castro, Kehagias}.
	
\end{itemize}

In a subsequent study, Ref.~\cite{Freitas:2022whn} extended the application of condition~(\ref{condition1freitas}) to additional models and found that
\begin{itemize}
    \item \textbf{Vector Field}: 
    \begin{itemize}
        \item{\bf Higher-dimensional models:} In contrast to the four-dimensional case, the zero modes of gauge fields in higher-dimensional braneworlds can lead to convergent integrals \cite{Oda03,Oda06,Midodashvili,Oda05,Costa01,Costa02,Flachi,Choudhury:2006nj,Arkani-Hamed:1999wga}. Nevertheless, all of these constructions fail to satisfy the generalized consistency condition (\ref{condition1freitas}). As a consequence, any higher-dimensional model involving free gauge fields must be excluded.
        \item{\bf $d=4$ with scalar field couplings:} 
        Many constructions introduce a coupling of the vector field to a background function $G(y)$ \cite{Kehagias,Fu01,Landim01,Landim02,Landim03,Chumbes:2011zt}, with the action  
        \begin{equation}\label{GF^2}
        	S = -\frac{1}{4} \int d^4x \, dy \, G(y) \, F^{MN} F_{MN}.
        \end{equation}
        By analyzing the equations of motion in conjunction with the generalized consistency condition, one finds that the only allowed choice is $G(y) = e^{2A}$. This requirement uniquely fixes the free parameters in the models of Refs.~\cite{Kehagias,Fu01,Landim01,Landim02,Landim03}, while simultaneously constraining or excluding all the constructions proposed in Ref.~\cite{Chumbes:2011zt}.
        \item \textbf{Non-minimal coupling:} We now turn to gauge fields with non-minimal couplings to gravity. In Refs.~\cite{Alencar02,Zhao}, a coupling of the form $\lambda_1 R A^M A_M$ was introduced, while Ref.~\cite{Alencar03} considered $\lambda_2 R^{MN} A_M A_N$. Finally, both terms were combined in Ref.~\cite{Freitas} to give the action
        \begin{equation}\label{coupling}
        	S = -\frac{1}{2} \int d^d x \, dy \left[ \frac{1}{2} F^{MN} F_{MN} + \lambda_1 R A^M A_M + \lambda_2 R^{MN} A_M A_N \right].
        \end{equation}
        Using the equations of motion and the separation of variables $\mathcal{A}_\mu(x,y) = \hat{\mathcal{A}}_\mu^T(x) \psi(y)$, the zero-mode solution is found to be $\psi(y) = e^{a A(y)}$, with $a = -2 \lambda_1 (D-1) - \lambda_2$, and satisfying
        \begin{equation}\label{l1l2}
        	\left[ \frac{D-4}{2} + a \right]^2 = \frac{(D-4)^2}{4} - \left( \lambda_1 (D-1) + \lambda_2 \right) (D-2).
        \end{equation}
        To ensure a normalizable zero mode, the condition $D-4 + 2a > 1$ must hold. Models with only one of the couplings, $\lambda_1=0$ or $\lambda_2=0$, leave no free parameters when condition~(\ref{l1l2}) is imposed. The first case was explored in Refs.~\cite{Alencar02,Zhao,Alencar:2015rtc}, and the second in Ref.~\cite{Alencar03}. However, the application of the full consistency condition shows that both couplings are required simultaneously. Consequently, all the models in Refs.~\cite{Alencar02,Zhao,Alencar:2015rtc,Alencar03} must be excluded. Furthermore, the consistency condition fully fixes the parameters in Ref.~\cite{Freitas} to
        \begin{equation}
        	\lambda_1 = -\frac{1}{(D-2)(D-1)}, \qquad \lambda_2 = -\frac{D-4}{D-2}.
        \end{equation}
    \end{itemize}
\end{itemize}

Therefore, only a limited number of models survive the application of the consistency condition. In the next subsection, we will show that none of the previous consistency conditions fully account for the coupled dynamics of gravity and bulk fields.

\subsection{Alencar \emph{et al.} Local Sum Rules}

	To fully capture the effects of the coupling between gravity and bulk fields, the authors of Ref.~\cite{Alencar:2024lrl} considered the five-dimensional metric  
\begin{equation}
	ds^2 = e^{2\sigma(y)} \hat{g}_{\alpha\beta}(x^\mu) dx^\alpha dx^\beta + dy^2,
\end{equation}
where Greek indices $\alpha, \beta$ run over the four-dimensional brane coordinates and $y$ denotes the extra-dimensional coordinate. By combining the five-dimensional Einstein tensor with the corresponding Einstein equations, they derived a tensorial relation that governs the coupled system:  
\begin{equation}\label{geova}
	(1 - m) \, {}^{(4)}G_{\mu\nu} = 8 \pi G_5 (1 - m) T_{\mu\nu} + 2 \pi (m - 2) G_5 e^{2\sigma} \big(T_\alpha^\alpha - 2 T_5^5 \big) \hat{g}_{\mu\nu} + \big( 3 \sigma'' + m \sigma'^2 \big) e^{2\sigma} \hat{g}_{\mu\nu}.
\end{equation}

	Contracting Eq.~(\ref{geova}) with $\hat{g}^{\mu\nu}$ reproduces the result of Gibbons \emph{et al.}, Eq.~(\ref{sumrule}). It is important to note, however, that Gibbons \emph{et al.} performed an integration over the extra dimension, which removes the last term in Eq.~(\ref{geova}) and leads to Eq.~(\ref{GSR}). Relying solely on this integrated condition (\ref{GSR}) has certain limitations. In particular, the integral of the last term in Eq.~(\ref{geova}) (or equivalently the left-hand side of Eq.~(\ref{sumrule})) vanishes only for the type I RS model, since the contributions from the two delta functions cancel. Therefore, the Gibbons \emph{et al.} condition is strictly valid for Type I braneworlds. Moreover, integrating over the extra dimension suppresses local features that may be relevant to field localization.

	Equation~(\ref{geova}) also recovers the Duff \emph{et al.} consistency condition when applied to the RS II model. Setting $m=0$, the last two terms cancel for $p$-form fields, reproducing their result. Importantly, the authors of Ref.~\cite{Alencar:2024lrl} considered a fully general energy-momentum tensor and an arbitrary warp factor, providing a broader generalization. This also extends the findings of Freitas \emph{et al.} \cite{Freitas}, who applied condition~(\ref{condition1freitas}) together with the field equations. In contrast, the local approach of Ref.~\cite{Alencar:2024lrl} shows that the equations of motion are not necessary, and the constraints can be expressed solely in terms of the bulk energy--momentum tensor.

	Additionally, it was demonstrated that any consistent background solution must satisfy
\begin{align}
	\label{geovabg} 
	\sigma'' + \sigma'^2 &= \frac{2 \pi G_5}{3} \left( T_\mu{}^\mu - 2 T_5{}^5 \right),\\
	\Lambda_b \, \hat{g}_{\mu\nu} &= 8 \pi G_5 T_{\mu\nu} - \pi G_5 e^{2\sigma} \left( 2 T_\alpha{}^\alpha - 4 T_5{}^5 \right) \hat{g}_{\mu\nu} - 3 \sigma'^2 e^{2\sigma} \hat{g}_{\mu\nu},
\end{align}
where the effective brane cosmological constant is defined as
\begin{equation}
	\Lambda_b = 4 \pi G_5 e^{2\sigma} T_5^5 + 3 \sigma'^2 e^{2\sigma}.
\end{equation}

It is straightforward to verify that the conditions above reproduce both the RS I and RS II models. In these cases, the energy-momentum tensor takes the form
\begin{equation}
T_{MN}=-g_{MN}\Lambda - \lambda_1\delta^{\mu}_{M}\delta^{\nu}_{N}\hat{g}_{\mu\nu} \delta(\phi-\pi)-\lambda_2 \delta(\phi)\delta^{\mu}_{M}\delta^{\nu}_{N}\hat{g}_{\mu\nu}.
\end{equation}
With the above expression, $\sigma=\kappa|y|$  and Eq. (\ref{geovabg}) we get
\[
-\kappa\delta(y)+\kappa\delta(y-L)+\kappa^{2}=-\frac{8\pi G_{5}}{3}\sum_{\alpha}\lambda_{\alpha}\delta(y-y_{\alpha})-\frac{4\pi G_{5}}{3}\Lambda.
\]

	Naturally, integrating both sides of the previous relation reproduces the standard fine-tuning condition $\lambda_1 = -\lambda_2$. However, expressing the constraint in its local form provides additional information. In particular, for $y \neq 0, \pi$ we find that the bulk cosmological constant must satisfy $\Lambda = -3 \kappa^2 / (8 \pi G_5)$. For the Type II scenario, consistency requires that $\lambda_2 = 0$.

When additional field sources are present in the bulk alongside the background-generating sources, the authors derived the following local constraints on the bulk matter fields:  
\begin{eqnarray}
	{}^{(b)}T_{\mu 5}(x^\alpha, y) &=& 0, \label{condition_1}\\
	{}^{(b)}T_\mu^\mu - 2 T_5^5 &=& 0, \label{geovatraco}\\
	{}^{(b)}T_{\mu\nu}(x^\alpha, y) &=& {}^{(b)}T_{\mu\nu}(x^\alpha), \label{geovamunu}\\
	{}^{(b)}T_{55}(x^\alpha, y) &=& -\frac{1}{16 \pi G_5} e^{-2\sigma} f(x^\alpha), \label{geova55}
\end{eqnarray}
where $f$ is an arbitrary function of the four-dimensional brane coordinates $x^\alpha$.

 Based on this, the authors applied their conditions to a series of fields and models, with the most important ones listed below:
 \begin{itemize}
     \item \textbf{Scalar fields}: 
     Considering a bulk scalar field with Lagrangian  
     \begin{equation}
     	\mathcal{L}(x^\alpha, y) = \frac{1}{2} (\partial \Phi)^2 + V(\Phi),
     \end{equation}
     and performing a separation of variables of the form $\Phi(x^\alpha, y) = \xi(y) \, \psi(x^\alpha)$, the application of the local consistency conditions leads to the solution $\xi(y) \sim \text{constant}$ and requires the potential to vanish, $V(\Phi) = 0$. This result is in agreement with the well-known conclusion that only free scalar fields can be consistently localized on the brane.

     \item \textbf{Fermion fields}: The authors then analyzed five-dimensional fermions, whose energy-momentum tensor is given by  
     \begin{equation}
     	T_{MN} = \frac{i}{2} \, \bar{\Psi} \, \Gamma_{(M} \nabla_{N)} \Psi - \frac{i}{2} \, \nabla_{(N} \bar{\Psi} \, \Gamma_{M)} \Psi.
     \end{equation}
     Applying a separation of variables $\Psi(x^\alpha, y) = \xi(y) \, \psi(x^\alpha)$, they derived the corresponding components of the energy--momentum tensor:
\begin{align}
T_{5\mu} &= \frac{i}{2}\,\xi^2 \left[ \bar\psi \gamma_5 \hat{\nabla}_\mu \psi - \hat{\nabla}_\mu \bar\psi\, \gamma_5 \psi \right] \bm{- \frac{i}{2} e^{-A} A' \xi^2 \bar\psi \gamma_\mu \psi} \label{condition_11}, \\[0.2cm]
T_{\mu\nu} &= \frac{i}{2} e^{\pmb{-}A} \xi^2 \left[ \bar\psi \gamma_{(\mu} \hat{\nabla}_{\nu)} \psi - \hat{\nabla}_{(\mu} \bar\psi\, \gamma_{\nu)} \psi \right]\bm{ - 2 i e^{2A} A' \xi^2 g_{\mu\nu} \bar\psi \psi} , \\[0.2cm]
T_\mu{}^{\mu} - T_5{}^{5} &= i \xi^2 e^{-\pmb{3}A} \left[ \bar\psi \gamma^\mu \hat{\nabla}_\mu \psi - \hat{\nabla}_\mu \bar\psi\, \gamma^\mu \psi \right] \bm{- 8 i A' \xi^2 \bar\psi \psi} , \\[0.2cm]
T_{55} &= 0.
\end{align}

Upon reexamining the analysis, we observed that the terms previously highlighted in bold are in fact zero, in addition to noting some necessary sign corrections and the presence of additional terms. We emphasize that, even after removing the vanishing contributions and implementing these corrections, the local consistency conditions remain unsatisfied. Consequently, the main conclusion of the original work, i.e., that free bulk fermion fields cannot be consistently localized, remains valid. For clarity, we point out that, contrary to the original claim, it is Eq.~\eqref{condition_11} that fails to satisfy the consistency conditions, since Eq.~\eqref{condition_1} could only be satisfied in a trivial scenario.

Thus, the correct components of the energy-momentum tensor are:
\begin{align}
T_{5\mu} &= \frac{i}{2}\,\xi^2 \left[ \bar\psi \gamma_5 \hat{\nabla}_\mu \psi - \hat{\nabla}_\mu \bar\psi\, \gamma_5 \psi \right] \label{condition_111}, \\[0.2cm]
T_{\mu\nu} &= \frac{i}{2} e^{A} \xi^2 \left[ \bar\psi \gamma_{(\mu} \hat{\nabla}_{\nu)} \psi - \hat{\nabla}_{(\mu} \bar\psi\, \gamma_{\nu)} \psi \right], \\[0.2cm]
T_\mu{}^{\mu} - 2T_5{}^{5} &= \frac{i}{2} \xi^2 e^{-A} \left[ \bar\psi \gamma^\mu \hat{\nabla}_\mu \psi - \hat{\nabla}_\mu \bar\psi\, \gamma^\mu \psi \right]  , \\[0.2cm]
T_{55} &= 0.
\end{align}
Regarding the Yukawa-type coupling, where the term 
$\lambda F(y)\bar{\Psi}\Psi$ is introduced into the action, 
which leads to the following contribution to the energy-momentum tensor:
\begin{equation}
\Delta T_{MN}(x^{\nu},y) 
= -\frac{1}{2}\, g_{MN}\, \lambda\, F(y)\, \bar{\Psi}\Psi,
\end{equation}
we emphasize that, contrary to the equation presented by the authors, 
the equation that fails to satisfy the consistency relations is equation \eqref{condition_111}. 
Moreover, adding this term does not change the outcome, since we still have 
$g_{5\mu}=0$, which means that models considering this type of coupling 
must also be ruled out.

     \item \textbf{Gauge fields}: Finally, the authors considered the consequences for the gauge field, whose energy-momentum tensor can be written as
     \begin{eqnarray}
          T_{MN}(x,y) = F_{MQ}F_{N}^Q - \frac{1}{4}g_{MN} F_{PQ}F^{PQ},
     \end{eqnarray}
     where $F_{MN} = \partial_M B_N - \partial_N B_M$ is the field strength tensor. Written $B_\mu$ in the form $B_\mu = \xi_B(y)\mathcal{B}(x^\alpha)$, the authors applied their conditions and found that
     \begin{eqnarray}
         T_\mu^\mu - 2T_5^5 = \frac{1}{2}e^{-4\sigma}\tilde{F}_{\mu\nu}\tilde{F}^{\mu\nu},
     \end{eqnarray}
     which violates condition \eqref{geovatraco}. Therefore, the gauge field is also not localized, which is indeed already known from the literature. In addition, the authors considered coupling models, with terms in the action of the form $G(y)F_{MN}F^{MN}$ or $\lambda_1RB^MB_M + \lambda_2R^{MN}B_MB_N$, where they showed that these coupling models also do not satisfy the presented consistency conditions.
 \end{itemize}

\subsection{Generalization of Gibbons \emph{et al.} consistency condition in arbitrary dimensions}

Up to this point, most of the consistency conditions presented have been limited to the case where the bulk has only five dimensions, or alternatively, to spacetimes with six dimensions in the bulk (codimension 2). However, in Ref. \cite{Leblond:2001xr}, the authors found a generalization of the Gibbons \emph{et al.} condition in a spacetime whose bulk has an arbitrary dimension. To do so, the authors started with a braneworld in $D = d + n$-dimensions, where the metric is given by the line element
\begin{equation}\label{metric}
    ds^2 = g_{MN}dx^Mdx^N = e^{2\sigma(y)}\hat{g}_{\mu\nu}(x)dx^\mu dx^\nu + \tilde{g}_{jk}(y)dy^jdy^k,
\end{equation}
where $d$ is the brane dimension, indexed by greek letters ($\mu,\nu,...$), and $n$ is the number of space-like extra dimensions, labeled by latin letters ( $j,k,...$).

For the metric \eqref{metric}, the components of the $D$-Einstein tensor and $D$-Ricci tensor are given by \cite{Freitas:2020mxr,Leblond:2001xr}
\begin{eqnarray}
       \label{teste} G_{\mu\nu} &=& ^{(d)}G_{\mu\nu}(x) + \hat{g}_{\mu\nu}(x)e^{2\sigma}\left[(d - 1)\nabla^2\sigma + \frac{d(d -1)}{2}(\nabla\sigma)^2 - \frac{^{(n)}R(y)}{2}\right],\\
\label{nondiagonal}            G_{\mu j} &=& 0,\\
\label{Gn}             G_{jk} &=& ^{(n)}G_{jk}(y) - \frac{1}{2}\tilde{g}_{jk}(y)\,^{(d)}R(x)e^{-2\sigma} - d(\nabla_j\nabla_k\sigma + \nabla_j\sigma\nabla_k\sigma)
      + \tilde{g}_{jk}(y)\left[d\nabla^2\sigma + \frac{d(d + 1)}{2}(\nabla\sigma)^2\right], \\
      R_{jk} &=& ^{(n)}R_{jk}(y) - d(\nabla_j\nabla_k\sigma + \nabla_j\sigma\nabla_k\sigma),\\
       R_{\mu\nu} &=& ^{(d)}R_{\mu\nu} - \hat{g}_{\mu\nu}(x)e^{2\sigma}(d(\nabla\sigma)^2 + \nabla^2\sigma),
\end{eqnarray}
where $(\nabla\sigma)^2 = \nabla_k\sigma\nabla^k\sigma$ and $\nabla^2\sigma = \nabla_k\nabla^k\sigma$.
 Taking the partial traces of the equations for the Ricci tensors, we have, respectively
 \begin{eqnarray}
   \label{ricci1}  \frac{1}{d}(^{(n)}R - R^j_j) &=& \nabla^2\sigma + (\nabla\sigma)^2,\\
  \label{ricci2}   \frac{1}{d}(^{(d)}R\,e^{-2\sigma} - R^\mu_\mu) &=& d(\nabla\sigma)^2 + \nabla^2\sigma,
 \end{eqnarray}
where the Ricci scalars are given by
\begin{eqnarray}
    R = R_{MN}g^{MN}, \qquad ^{(d)}R =\; ^{(d)}R_{\mu\nu}\hat{g}^{\mu\nu}, \qquad ^{(n)}R =\; ^{(n)}R_{jk}\tilde{g}^{jk}.
\end{eqnarray}

For convenience, we rewrite the Eqs.~\eqref{ricci1} and \eqref{ricci2} in the following form
\begin{align}
	\label{teste2} 
	\frac{1}{d} \left( {}^{(n)}R - R^j_j \right) &= e^{-\sigma} \nabla^2 \big( e^{\sigma} \big),\\
	\frac{1}{d} \left( {}^{(d)}R \, e^{-2\sigma} - R^\mu_\mu \right) &= (d - 1) e^{-2\sigma} \big( \nabla e^{\sigma} \big)^2 + e^{-\sigma} \nabla^2 \big( e^{\sigma} \big).
\end{align}

Moreover, it is straightforward to arrive at the following total derivative identity in this internal space
\begin{eqnarray}
    \nabla \cdot (e^{\alpha\sigma}\nabla e^\sigma) = e^{(\alpha + 1)\sigma}(\alpha e^{-2\sigma}(\nabla e^\sigma)^2  + e^{-\sigma} \nabla^2 e^\sigma),
\end{eqnarray}
where $\alpha$ is a constant. 
Using this identity and the equations for the Ricci scalar, we have
\begin{eqnarray}\label{ricci3}
       \nabla \cdot (e^{\alpha\sigma}\nabla e^\sigma) = \frac{e^{(\alpha + 1)\sigma}}{d(d - 1)}[\alpha(^{(d)}Re^{-2\sigma} - R^\mu_\mu) + (d - 1 - \alpha)(^{(n)}R - R^j_j)].
\end{eqnarray}

Consider the Einstein equation in $D = d + n$ dimensions,
\begin{equation}
    R_{MN} - \frac{1}{2}g_{MN}R = 8\pi G_{D}T_{MN},
\end{equation}
where $G_D$ denotes the Newton constant in $D = d + n$ dimensions, this can be rewritten as
\begin{eqnarray}
    R_{MN} = 8\pi G_D\left(T_{MN} - \frac{1}{d + n - 2}g_{MN}T^P_P\right),
\end{eqnarray}
where $T^P_P = T^\mu_\mu + T^j_j$ and $T^\mu_\mu = e^{-2\sigma}T_{\mu\nu}\hat{g}^{\mu\nu}$.
We can write the traces of the Ricci tensors as
\begin{align}
	R^\mu_\mu &= \frac{8 \pi G_D}{d + n - 2} \Big[ (n - 2) T^\mu_\mu - d \, T^j_j \Big],\\
	\label{teste5}
	R^j_j &= \frac{8 \pi G_D}{d + n - 2} \Big[ (d - 2) T^j_j - n \, T^\mu_\mu \Big].
\end{align}

Substituting this into Eq.~\eqref{ricci3}, we have
\begin{eqnarray}
  \label{leblondc}   \nabla \cdot (e^{\alpha\sigma}\nabla e^\sigma) &=& \frac{e^{(\alpha + 1)\sigma}}{d(d - 1)}\left\{\frac{8\pi G_D}{d + n - 2}\left(T^\mu_\mu[\alpha( n - 2) + n( d - 1 - \alpha)]\right.\right. \nonumber \\
     &&+ \left.\left.T^j_j[\alpha d  - (d - 2)(d - 1 - \alpha)]\right)
     + (d - 1 - \alpha)^{(n)}R + \alpha ^{(d)}Re^{-2\sigma}\right\}.
\end{eqnarray}

In a compact internal space, the integral of the total derivative on the LHS is zero, which yields
\begin{eqnarray}
\label{condition1leblon}    \oint e^{(\alpha + 1)\sigma}\left(T^\mu_\mu[\alpha(n - 2) + n(d - 1 - \alpha)] + T^j_j[\alpha d - (d - 2)(d - 1 - \alpha))]\right.\\\nonumber 
    +\left. \frac{d + n - 2}{8\pi G_D}[(d - 1 - \alpha)^{(n)}R + \alpha\; ^{(d)}Re^{-2\sigma}]\right) = 0.
\end{eqnarray}

For $\alpha = m - 1$, $d = 4$, $n = 1 (^{(n)}R = 0)$, the equation above reduces to the consistency condition of Gibbons \emph{et al.} \cite{Gibbons:2000tf}. From this, the authors considered an energy-momentum tensor of the form
\begin{eqnarray} \label{Tvacuogeral} 
  T_{MN} = -\Lambda g_{MN} - \sum_i T_q^{(i)}P[g_{MN}]_q^{(i)}\Delta^{(D - q - 1)}(y - y_i) + \tau_{MN},
\end{eqnarray}
where $\Lambda$ is a bulk cosmological constant, and the other term describes a collection of branes of various dimensions, that is, a collection of  $q-$branes ($q \geq d - 1$) of various dimensions with tension $T_q^{(i)}$ and transverse coordinates $y_i$. Furthermore, $P[g_{MN}]_q^{(i)}$ is the pull-back of the spacetime metric to the worldvolume of the $q$-brane \cite{Leblond:2001xr} and $\Delta^{(D - q - 1)}(y - y_i)$ are  a combination of covariant deltas that position the brane in $y_i$. Futhermore, any other bulk or worldvolume matter field contributions is in $\tau_{MN}$.

As a first application example, the authors considered the simple case where $d = 4$, $n = 1$, $q = 3$, $\alpha = - 1$ and $\tau_{MN} = 0$, where the condition reduces to
\begin{eqnarray}
    -^{(d)}R\oint e^{-2\sigma} = 32\pi G_5 \sum_i T_3^{(i)}.
\end{eqnarray}

This condition was first obtained in \cite{Gibbons:2000tf}. If 
$^{(d)}R \geq 0$, then $\sum_i T_3^{(i)} < 0$, and this imposes the need to introduce negative tension branes for consistency. Moreover, as stated in \cite{Leblond:2001xr}, since $\tau_{MN} = 0$, $^{(d)}R$ must be constant. This statement can be demonstrated later when we introduce our local consistency conditions.

Another interesting example considered is a six-dimensional spacetime,  and in which the tensor $\tau_{MN}$ is determined by a gauge field whose field strength $F_{ij}$ is proportional to the volume form of the two-dimensional internal space, that it's
\begin{eqnarray}
    F_{ij} = k \epsilon_{ij},
\end{eqnarray}
where $k$ is the proportionality constant. With this, condition \eqref{leblondc} reduces to
\begin{eqnarray}
    \oint e^{2(\alpha + 1)\sigma}\left(\alpha\; ^{(d)}Re^{-2\sigma} + (3 - \alpha)\;^{(n)}R - (3 + \alpha)\Lambda - 2\pi G_6(9 - 5\alpha k^2\right)\\\nonumber
    = 4\pi G_6\left(4(3 - \alpha)\sum_i T_3^{(i)}e^{2(\alpha + 1)\sigma(y_i)} + (9 - \alpha)\sum_i T_4^{(i)}\oint_i e^{2(\alpha + 1)\sigma}\right),
\end{eqnarray}
where $\oint_i e^{2(\alpha + 1)\sigma}$ is the integral over the one-cycle spanned by the $i^{\text{th}}$ four-brane in the internal space.
When we consider only 3-branes and a non-warped spacetime ($\sigma = 0$), and furthermore $k = \Lambda = ^{(d)}R$, we recover the condition given by
\begin{eqnarray}\label{xiref}
    \chi = 4G_6\sum_i T_3^{(i)},
\end{eqnarray}
where $\chi$ is defined as
\begin{eqnarray}
    \chi = \frac{1}{4\pi}\oint ^{(d)}R.
\end{eqnarray}
	Here, $\chi$ denotes the Euler--Poincaré characteristic, a topological invariant that characterizes the shape of the internal space and depends on the specific braneworld model under consideration. For instance, if $\chi > 0$, as in the case of a spherical internal space ($\chi = 2$), it is possible to construct a braneworld configuration with only positive-tension 3-branes. On the other hand, if $\chi = 0$, as in Ref.~\cite{Arkani-Hamed:1998sfv}, the introduction of negative-tension branes becomes necessary to satisfy Eq.~\eqref{xiref}.

Still in the six-dimensional case, the authors considered the more general scenario where $k \neq 0$ and $\Lambda \neq 0$. In this case, by combining the conditions for different values of $\alpha$, it is possible to arrive at the following conditions
\begin{eqnarray}
    \Lambda &=& \frac{k^2}{4},\\
    \chi - 2G_6V_2k^2 &=& 4G_6\sum_iT_3^{(i)},
\end{eqnarray}
where $V_2$ is the volume of the internal spcae. The introduction of these extra parameters makes it more challenging to construct consistent compactifications with only positive tension branes \cite{Leblond:2001xr}. In summary, one of the most notable outcomes derived from the consistency conditions presented in \cite{Leblond:2001xr} is the requirement for negative tension branes to appear in five-dimensional scenarios. However, the authors demonstrate how this result can be avoided in braneworld models with more than five dimensions. Another interesting point is that, since the authors generalized the condition of Gibbons \emph{et al.}, their conditions, although applicable to arbitrary dimensions, remain global conditions that do not account for the local nature of the Einstein equations.

\section{Local sum rules for bulk fields in arbitrary dimensions}\label{SecIII}

The generalization of the Gibbons \emph{et al.} condition presented in Ref.~\cite{Leblond:2001xr}, while providing an important extension that examines consistency conditions in a more general framework and has notable implications as discussed previously, still does not fully incorporate the complete dynamics arising from the coupling between gravity and bulk fields in arbitrary dimensions.

Therefore, in this section we derive four local consistency conditions for braneworld scenarios in an arbitrary number of dimensions, providing a generalization of all previously known consistency conditions. Our metric in arbitrary dimensions is given in \eqref{metric}. For convenience, we rewrite Eqs.~\eqref{teste} and \eqref{teste2} in the following form:
\begin{eqnarray}
   \label{teste3} G_{\mu\nu} &=& ^{(d)}G_{\mu\nu} + \hat{g}_{\mu\nu}e^{2\sigma}\left[(d - 1)\nabla^2\sigma + \frac{d(d - 1)}{2}(\nabla\sigma)^2\right] - \hat{g}_{\mu\nu}e^{2\sigma}\frac{^{(n)}R}{2},\\
   \label{teste4} R^j_j &=& ^{(n)}R - d(\nabla^2\sigma + (\nabla\sigma)^2).
\end{eqnarray}

By multiplying Eq.~\eqref{teste3} by $\gamma_1\,e^{(m - 2)\sigma}$ and Eq.~\eqref{teste4} by $\gamma_2\,e^{m\sigma}\hat{g}_{\mu\nu}$
, and subsequently summing them, we obtain
\begin{eqnarray}
    &&\frac{2(m - 1)}{d - 2}e^{(m - 2)\sigma}[G^{(D)}_{\mu\nu} - G^{(d)}_{\mu\nu}] + \left(\frac{d - 1}{d}\right)\left[\frac{2(m - 1)}{d - 2} - 1\right]e^{m\sigma}\hat{g}_{\mu\nu}R^j_j
    	\nonumber \\
    && \qquad = (d - 1)\nabla\cdot(e^{m\sigma}\nabla\sigma)\hat{g}_{\mu\nu} 
    + R^{(n)}e^{m\sigma}\left\{\left(\frac{d - 1}{d}\right)\left[\frac{2(m - 1)}{d - 2} - 1\right] - \frac{1}{2}\frac{2(m - 1)}{d - 2}\right\}\hat{g}_{\mu\nu},
\end{eqnarray}
where
\begin{eqnarray}
    \gamma_1 &=& \frac{2(m - 1)}{d - 2},\\
    \gamma_2 &=& \frac{(d - 1)}{d}\left[\frac{2(m - 1)}{d - 2} - 1\right],
\end{eqnarray}
and $m$ is a constant.

Now, using the $D - $dimensional Einstein equation
\begin{equation}
    G_{\mu\nu} = 8\pi G_DT_{\mu\nu},
\end{equation}
and Eq.~\eqref{teste5}, we have
\begin{align}
	\label{expressao}
	\frac{2(1 - m)}{d - 2} \, {}^{(d)}G_{\mu\nu} &= \frac{16 \pi G_D (1 - m)}{d - 2} \, T_{\mu\nu} + (d - 1) \big( \nabla^2 \sigma + m (\nabla \sigma)^2 \big) e^{2\sigma} \hat{g}_{\mu\nu} \nonumber \\
	&\quad + \frac{8 \pi (d - 1) G_D}{d (d + n - 2)} \left[ \frac{2 (m - 1)}{d - 2} - 1 \right] e^{2\sigma} \big[ n T^\alpha_\alpha - (d - 2) T^j_j \big] \hat{g}_{\mu\nu} \nonumber \\
	&\quad + {}^{(n)}R \, e^{2\sigma} \left( \frac{m - d}{d} \right) \hat{g}_{\mu\nu}.
\end{align}

	This local tensor equation provides a unified generalization of all previously discussed consistency conditions for bulk fields. For example, contracting Eq.~\eqref{expressao} with the metric $\hat{g}^{\mu\nu}$ and integrating over the internal space reproduces the result of Ref.~\cite{Leblond:2001xr}, Eq.~\eqref{condition1leblon}, upon identifying $\alpha = m - 1$, thereby generalizing the Gibbons condition. In the special case $d = 4$ and $n = 1$, performing the integration recovers the original Gibbons condition \cite{Gibbons:2000tf}. Similarly, setting $d = 4$ and $n = 1$ without integration reproduces the equation derived in Ref.~\cite{Alencar:2024lrl}, which represents the most general local consistency condition for 3-branes with a single codimension. Before discussing the consequences for bulk fields, it is essential to first analyze the background (vacuum) solution of the setup.

\subsection{Background solution}

We now turn to the analysis of the vacuum solution. To this end, we set $m = 1$ in Eq.~\eqref{expressao}, which corresponds to focusing on the background geometry without contributions from bulk matter fields. This choice simplifies the equation significantly, isolating the terms that depend solely on the warp factor and the curvature of the internal space. Explicitly, we obtain
\begin{eqnarray}
	\label{nabla1} \nabla^2\sigma + (\nabla\sigma)^2 = \frac{8\pi G_D}{d(d + n - 2)}[n T^\alpha_\alpha - (d - 2)T^j_j] + \, \frac{^{(n)}R }{d}.
\end{eqnarray}
This expression governs the warp factor $\sigma(y)$ and the internal space curvature in the absence of bulk fields, providing the foundation upon which localized matter fields can be consistently analyzed.

Now, setting $m = 0$ in Eq.~\eqref{expressao} and using the previous result, we obtain
\begin{eqnarray}
\label{dG}    ^{(d)}G _{\mu\nu} &=& 8\pi G_DT_{\mu\nu} - \frac{8\pi G_D(d - 1)}{d(d + n - 2)}[n T^\alpha_\alpha - (d - 2)T^j_j]e^{2\sigma}\hat{g}_{\mu\nu} \nonumber \\
    && - \frac{(d - 1)(d - 2)}{2}e^{2\sigma}(\nabla\sigma)^2 \hat{g}_{\mu\nu} - \frac{ d - 2}{2d}{}^{(n)}Re^{2\sigma}\hat{g}_{\mu\nu}.
\end{eqnarray}

We assume that the background is produced by an independent source, denoted by $^{(v)}T_{MN}(y)$, representing the vacuum contribution. Under this assumption, and taking the brane to have constant curvature, the background solution is determined from
\begin{eqnarray}
\label{lambdaeq}        
\Lambda_b\hat{g}_{\mu\nu} &=& 8\pi G_D {}^{(v)}T_{\mu\nu}(y) - \frac{8\pi G_D(d - 1)}{d(d + n - 2)}[n\, ^{(v)}T^\alpha_\alpha (y) - (d - 2)\, ^{(v)}T^j_j(y)]e^{2\sigma}\hat{g}_{\mu\nu} 
    \nonumber \\
    && - \frac{(d - 1)(d - 2)}{2}e^{2\sigma}(\nabla\sigma)^2 \hat{g}_{\mu\nu} - \frac{ d - 2}{2d}   \,{}^{(n)}R(y)e^{2\sigma}\hat{g}_{\mu\nu},\\
          \label{nablaeq}  \nabla^2\sigma + (\nabla\sigma)^2 &=& \frac{8\pi G_D}{d(d + n - 2)}[n\,  ^{(v)}T^\alpha_\alpha (y) - (d - 2)\, ^{(v)}T^j_j(y)] + {}^{(n)}R(y)\frac{1}{d},
\end{eqnarray}
where
\begin{eqnarray}
    \Lambda_b &=& \frac{8\pi G_D( d - 2)(1 - n)}{d(d + n - 2)}e^{2\sigma}T^\alpha_\alpha + \frac{8\pi G_D (d - 1)(d - 2)}{d(d + n - 2)}e^{2\sigma}T^j_j
        \nonumber \\
    && - \frac{(d - 1)(d - 2)}{2}e^{2\sigma}(\nabla\sigma)^2 - {}^{(n)}R\frac{d - 2}{2d}e^{2\sigma}.
\end{eqnarray}
	A careful examination of the above equations reveals some differences from the $d = 4$, $n = 1$ case. In the present setup, the background solution explicitly depends on the Ricci scalar of the internal space, ${}^{(n)}R$, which is natural since the internal space is no longer assumed to be flat. This dependence is trivial in the codimension-one case ($n = 1$), where the internal space is necessarily one-dimensional and flat.

Now, following Ref.~\cite{Leblond:2001xr}, we consider the same ansatz for the energy-momentum tensor, given by Eq.~\eqref{Tvacuogeral}, where we have the traces given by
\begin{eqnarray}
    ^{(v)}T^\mu_\mu &=& -d\left[\Lambda + \sum_i T_q^{(i)}\Delta^{(D - q -1)}(y - y_i)\right]\\
    ^{(v)}T^j_j &=& - n\Lambda - \sum_i (q - d + 1)T_q^{(i)}\Delta^{(D - q - 1)}(y - y_i).
\end{eqnarray}

Substituting this into the Eq.~\eqref{nablaeq}, we have
\begin{eqnarray}
 \label{vacum}   \nabla^2\sigma + (\nabla\sigma)^2 &=& \frac{8\pi G_D}{d(d + n - 2)}\sum_i [(d - 2)(q - d + 1) - nd]T_q^{(i)}\Delta^{(D - q - 1)}(y - y_i)\nonumber \\
    &&-  \frac{16\pi G_D n \Lambda}{d(d + n - 2)} + R^{(n)}(y)\frac{1}{d}.
\end{eqnarray}

The consistency equation presented establishes a general relationship between the braneworld geometry and the energy-momentum tensor. However, due to its complexity, analyzing this equation in its general form is challenging and heavily depends on the specific braneworld model considered. To illustrate its effectiveness, we will first apply this equation to the crucial case: the original Randall-Sundrum I and II models. In this context, $d = 4$, $n = 1$, $^{(1)}R = 0$ and $\sigma(y) = \kappa|y|$, and the equation above reduces to
\begin{eqnarray}
    -\kappa\delta(y) + \kappa\delta(y - y_1) +\kappa^2 = - \frac{4\pi G_5 \Lambda}{3} - \frac{8\pi G_5}{3}\sum_i T_3^{(i)}\delta(y - y_i)
.\end{eqnarray}
Integrating the equation we have the RS fine-tuning $T_3^{(1)} = - T_3^{(2)}$. If we have $y \neq 0, y_1$, then $\Lambda = - 3\kappa^2/8\pi G_5$, and the RS type II is recovered when $T_3^{(2)} = 0$.

Another interesting vacuum solution is given by a non-warped geometry $\sigma = 0$ in a $D = 6$ braneworld with $d = 4$, $n =  2$ and without a cosmological constant $\Lambda = 0$. Furthermore, if we consider $q = 3$, that is, only 3-branes, then Eq.~\eqref{vacum} becomes
\begin{eqnarray}
 \label{nonwarped}   R^{(2)} = 16\pi G_6\sum_i T_3^{(i)}\Delta^{(2)}(y - y_i),
\end{eqnarray}
where $R^{(2)}$ is the Ricci scalar of the internal space. This condition is local and expresses the relationship between the tensions on the branes and the curvature of the internal space. If we integrate Eq.~\eqref{nonwarped}, we get the same condition presented in Ref. \cite{Leblond:2001xr}
\begin{eqnarray}
  \label{nonwpaerd2}  \chi = 4 G_6\sum_i T_3^{(i)}.
\end{eqnarray}

In Eq.~\eqref{nonwarped}, for $y \neq y_i$, we have 
\begin{eqnarray}
    R^{(2)} = 0.
\end{eqnarray}

We can, still following Refs.~\cite{Leblond:2001ex,Leblond:2001xr}, introduce a gauge field that exists only in the internal space, characterized by a field strength that is proportional to the
volume form of the two-dimensional internal space. 
With this, we have that
\begin{eqnarray}
    \tau^\mu_\mu = -k^2\,, \qquad \tau_j^j = \frac{k^2}{2},
\end{eqnarray}
and using Eq.~\eqref{dG} with $m = 4$, $\sigma = 0$ and $G^{(d)}_{\mu\nu} = 0$, we have the condition
\begin{eqnarray}
  \label{lambdacondition}  \Lambda = \frac{k^2}{4} > 0.
\end{eqnarray}
On the other hand, Eq.~\eqref{nablaeq} provides
\begin{eqnarray}
   R^{(2)} - 2\pi G_6k^2 - 24\pi G_6\Lambda = 16\pi G_6 \sum_i T_3^{(i)}\Delta^{(2)}(y - y_i).
\end{eqnarray}

Using relation \eqref{lambdacondition}, we have
\begin{eqnarray}
  \label{conditionCD2}  R^{(2)} - 8\pi G_6 k^2 = 16\pi G_6 \sum_i T_3^{(i)}\Delta^{(2)}(y - y_i).
\end{eqnarray}

If we integrate, we have the global condition
\begin{eqnarray}
    \chi - 2G_6 V_2 k^2 = 4 G_6\sum_i T_3^{(i)},
\end{eqnarray}
which was obtained in \cite{Leblond:2001xr}. For $y \neq y_i$ in Eq.~\eqref{conditionCD2}, we have 
\begin{eqnarray}
    R^{(2)} = 8\pi G_6 k^2  = 32\pi G_6 \Lambda > 0.
\end{eqnarray}

It is noteworthy that the local relations expressed in Eqs.~\eqref{nonwarped} and \eqref{conditionCD2} already encode, in differential form, the same information contained in the global constraints derived in Ref.~\cite{Leblond:2001xr}. Indeed, integrating these expressions over the internal space reproduces the standard consistency conditions, thus recovering the Gibbons-type global constraints. However, the local formulation is not merely a sufficient mathematical convenience but a physical necessity for the consistency of the $D$-dimensional theory. While global constraints ensure that the effective 4D theory is well-defined on average, they can fail to detect local singularities or incompatible stress-energy distributions at specific points in the bulk. By describing, point by point, how each brane contributes to the curvature through the localized term $\Delta^{(D-q-1)}(y - y_i)$, our approach ensures that the gravitational backreaction is compatible with the warped-product symmetry everywhere. If these pointwise conditions were violated, the resulting backreaction would inevitably force the metric to deviate from the assumed ansatz--for instance, by inducing $x$-dependence in the warp factor or causing brane-bending, even if the integrated global condition remained satisfied. Thus, the local consistency conditions provide a more stringent and complete framework, ensuring that the braneworld geometry is microscopically stable and consistent with the full set of Einstein equations.

We emphasize that, for codimension greater than one, thin-brane sources may lead to singular transverse geometries. In particular, codimension-two branes are commonly associated with conical singularities, so that the curvature may contain localized distributional contributions at the brane position \cite{Geroch:1987qn,Traschen:2008cy,Vinet:2005dg,Maartens:2010ar}. In such cases, the pointwise form of the local Einstein equations must be interpreted with care and, at the singular locus, generally requires a model-dependent prescription, such as matching conditions, an excision or near-brane treatment, or a finite-thickness regularization \cite{Bostock:2003cv,Navarro:2004di,Papantonopoulos:2006uj}. Therefore, the local consistency conditions derived here should be understood as directly applicable to smooth regions of the internal space, as well as to smooth or regularized braneworld backgrounds where the metric, warp factor, and internal curvature are well-defined local quantities.

Now, in addition to the source that generates the background, we will consider the presence of matter fields in the bulk.

\subsection{Consequences for bulk fields}

We now extend our analysis to include, in addition to the vacuum sources, the presence of matter fields propagating in the bulk. It is important to emphasize that we do not treat the background geometry as a rigid manifold; rather, the metric components remain dynamic and are subject to the full backreaction of any additional matter content. By incorporating these bulk fields into the energy-momentum tensor, we investigate the dynamic interplay between matter and geometry through the warp factor and the internal space curvature. The resulting local consistency conditions should, therefore, be understood as integrability requirements dictated by the Einstein equations. They define the specific class of bulk sources whose backreaction is compatible with the existence of a warped-product spacetime given by the metric \eqref{metric}. This approach provides a rigorous framework for determining which field configurations can be consistently localized on the brane while allowing the geometry to respond dynamically to the bulk matter, ensuring that the overall gravitational structure remains self-consistent in arbitrary dimensions.

Thus, we write the energy-momentum tensor in the form
\begin{equation}
	\label{generalT}
	T_{MN} = {}^{(v)}T_{MN}(y) + {}^{(b)}T_{MN}(x, y),
\end{equation}
where $^{(b)}T_{MN}(x,y)$ is the energy momemtum tensor for the matter fields.
To obtain the consistency conditions for the matter fields, we assume that $^{(v)}T_{MN}(y)$ corresponds to the source responsible for generating the background geometry. In addition, we make use of the local tensorial equations derived in the previous section.

The first consistency condition is trivially obtained from Eq.~\eqref{nondiagonal}, that is
\begin{eqnarray}
   \label{condition1} ^{(b)}T_{\mu j}(x, y) = 0. 
\end{eqnarray}

The second condition is obtained by substituting Eq.~\eqref{generalT} into Eq.~\eqref{nabla1}, such that
\begin{eqnarray}
     \nabla^2\sigma (y) + \nabla\sigma (y)\cdot\nabla\sigma (y) - \frac{8\pi G_D}{d(d + n - 2)}[n\, ^{(v)}T^\alpha_\alpha (y)
     - (d - 2)^{(v)}T^j_j(y)] 
        \nonumber \\
     - \frac{8\pi G_D}{d(d + n - 2)}[n\, ^{(b)}T^\alpha_\alpha - (d - 2)^{(b)}T^j_j] - \frac{1}{d}{}^{(n)}R(y) = 0.
\end{eqnarray}

Considering the background solution obtained earlier, we have the condition
\begin{eqnarray}
\label{condition2}    n\, ^{(b)}T^\alpha_\alpha - (d - 2)^{(b)}T^j_j = 0.
\end{eqnarray}
For $n = 1$ and $d = 4$ we recover $^{(b)}T^\alpha_\alpha - 2^{(b)}T^5_5 = 0$ from \cite{Alencar:2024lrl}. Equation \eqref{condition2} is a generalization of the expression obtained in \cite{Alencar:2024lrl} for a arbritary branewold model.

The third condition is obtained by substituting Eqs. \eqref{lambdaeq}, \eqref{generalT} and \eqref{condition2} in Eq.~\eqref{dG}, where we obtain
\begin{eqnarray}
    ^{(d)}G_{\mu\nu} = 8\pi G_D ^{(b)}T_{\mu\nu} + \Lambda_b \hat{g}_{\mu\nu}.
\end{eqnarray}

Since the LHS of the above equation depends only on $x^\alpha$ , the RHS must also depend only on this coordinates, thus we have the condition
\begin{eqnarray}
 \label{condition3}   {}^{(b)}T_{\mu\nu} (x, y) =  {}^{(b)}T_{\mu\nu} (x),
\end{eqnarray}
which was obtained in Ref.~\cite{Freitas:2020mxr} and subsequently in \cite{Alencar:2024lrl}.

The last condition is obtained from Eq.~\eqref{Gn} using Einstein's equation and the background solution, from which we obtain 
\begin{eqnarray}
    ^{(b)}T_{jk}(x,y) = - \frac{1}{16\pi G_D}\tilde{g}_{jk}(y)e^{-2\sigma}\; {}^{(d)}R(x^\alpha),
\end{eqnarray}
which is the condition obtained by \cite{Freitas:2020mxr}. Writing it in the form
\begin{eqnarray}
 \label{condition4}   ^{(b)}T_j^j(x^\alpha, y^i) = - \frac{n}{16\pi G_D}e^{-2\sigma}f(x^\alpha), 
\end{eqnarray}
where $f$ is a function that depends only on brane coordinates. The last equation can easily be identified with the condition obtained in \cite{Alencar:2024lrl}.

We can thus summarize the local consistency conditions for bulk fields in braneworlds with an arbitrary number of dimensions as follows:
\begin{eqnarray}
   &&\;\;\; (\text{CCI}):\;\; ^{(b)}T_{\mu j}(x, y) = 0, \\
   &&\;\;\; (\text{CCII}):\;\; n\, ^{(b)}T^\alpha_\alpha - (d - 2)^{(b)}T^j_j = 0, \\
   && \;\;\; (\text{CCIII}):\;\;  ^{(b)}T_{\mu\nu} (x, y) =  \,{}^{(b)}T_{\mu\nu} (x),\\
   &&\;\;\; (\text{CCIV}):\;\;  ^{(b)}T_j^j(x, y) = - \frac{n}{16\pi G_D}e^{-2\sigma}f(x^\alpha)\,.
\end{eqnarray}

Next, we investigate the consequences of these conditions for the localization of the zero mode of various bulk fields.

\section{Aplications on bulk fields}\label{SecIV}

	In this section, we turn our attention to the implications of these local consistency conditions for the localization of zero modes of bulk fields. The zero mode is particularly important because it determines the effective low-energy behavior of the field on the brane, corresponding to the massless excitations that can propagate in four dimensions. By analyzing the constraints imposed by the local consistency relations, we can identify which types of fields, such as scalars, gauge fields, or fermions, can have normalizable zero modes confined to the brane, and under what circumstances. This analysis allows us to systematically determine the viability of different field configurations in arbitrary-dimensional braneworld scenarios and to understand how the interplay between bulk dynamics and the geometry of the extra dimensions affects the localization mechanisms.

\subsection{Scalar field}

As a first application of our conditions, we can consider a Lagrangian of the scalar field $\Phi(x,y)$
\begin{eqnarray}
    \mathcal{L}(x,y) = \frac{1}{2}(\partial\Phi)^2 + V(\Phi),
\end{eqnarray}
whose energy-momentum tensor is given by
\begin{eqnarray}
    ^{(b)}T_{MN}(x,y) = \partial_M\Phi(x,y)\partial_N\Phi(x,y) - g_{MN}\left(\frac{1}{2}(\partial\Phi)^2 - V(\Phi)\right).
\end{eqnarray}

If we use separation of variables to write the scalar field as $\Phi(x,y) = \xi (y^1,y^2,...,y^n)\phi(x^\alpha)$, and apply CCI, we have that
\begin{eqnarray}
    \xi(y^1,y^2,...,y^n) \sim \text{constant}.
\end{eqnarray}

This result is consisent with CCIII. If we apply CII, we have that
\begin{eqnarray}
    V(\Phi) = 0,
\end{eqnarray}
such that we conclude that only free fields are localized when considering braneworlds in arbitrary dimensions. An interesting aspect is that this result does not depend on the geometry of the internal space.

\subsection{Bosonic fields}

\begin{itemize}
    
\item \textbf{Vector field}: We will now consider a gauge field whose Lagrangian is given by
\begin{eqnarray}
    L(x,y) = - \frac{1}{4}F_{MN}F^{MN},
\end{eqnarray}
where $F_{MN} = \partial_M A_{N} - \partial_N A_M$ is the field strength tensor. The energy-momentum tensor is
\begin{eqnarray}
    ^{(b)}T_{MN}(x,y) = F_{MQ}F_{N}^Q - \frac{1}{4}g_{MN} F_{PQ}F^{PQ}.
\end{eqnarray}

The vector $A_M$ is represented by $A_M = (A^T_\mu, B_k)$. If we use separation of variables to write $A^T_\mu (x,y) = \xi(y^1,y^2,...,y^n)\hat{A}_\mu(x^\alpha)$, then the CCI
implies that
\begin{eqnarray}
  \xi(y^1,y^2,...,y^n) \sim \text{constant}.  
\end{eqnarray}
However, with this we have that
\begin{eqnarray}
    ^{(b)}T_{\mu\nu} = e^{-2\sigma}\left(\hat{F}_{\mu\alpha}\hat{F}_\nu^{\alpha} - \frac{\hat{g}_{\mu\nu}}{4}\hat{F}_{\alpha\beta}\hat{F}^{\alpha\beta}\right),
\end{eqnarray}
which contradicts the condition CCIII. Moreover, the condition CCII implies that
\begin{eqnarray}
    \frac{n e^{-4\sigma}\hat{F}_{\mu\nu}\hat{F}^{\mu\nu}}{2} = 0,
\end{eqnarray}
which would imply a null effective action. Thus, the gauge field does not satisfy our conditions and is therefore not localized.

\item \textbf{A no-go theorem for $p$-form fields}: Given an arbitrary free
$p$-form field $\mathcal{A}_{N_1N2...N_p}$, the only $p$-form that can be localized in a braneworld with \textbf{arbitrary dimensions and internal space}, while respecting the consistency conditions for fields in the bulk is the $0$-form, i.e., a scalar field.

\textbf{Proof.}: The action for the $p-$form is
\begin{eqnarray}
    S^{(b)} = -\frac{1}{2(p + 1)!}\int d^dxd^ny\sqrt{-g}\mathcal{F}_{M_1...M_{p + 1}}\mathcal{F}^{M_1...M_{p + 1}},
\end{eqnarray}
where $\mathcal{F}_{M_1...M_{p + 1}} = \partial_{[N_1}\mathcal{A}_{N_2...N_{p+1}]}$. The energy-momentum tensor for this field is given by
\begin{eqnarray}
    ^{(b)}T_{MN}(x^\alpha, y^i) = \frac{1}{p!}\mathcal{F}_{MM_2...M_{p + 1}}\mathcal{F}_N^{M_2...M_{p + 1}} - \frac{1}{2(p + 1)!}g_{MN}\mathcal{F}_{M_1...M_{p + 1}}\mathcal{F}^{M_1...M_{p + 1}}.
\end{eqnarray}

If we consider that only the components $\mathcal{A}_{\mu_1\mu_2...\mu_{p}}$ are nonzero and use variable separation to write $\mathcal{A}_{\mu_1\mu_2...\mu_{p}} = \hat{\mathcal{A}}_{\mu_1\mu_2...\mu_{p}}(x^\alpha)\xi_A(y^1,y^2,...,y^n)$, then the $\mu_1-j$ component of $T_{MN}$ are
\begin{eqnarray}
    ^{(b)}T_{\mu_1j} = \frac{1}{p!}\hat{\mathcal{F}}_{\mu_1\mu_2...\mu_{p+1}}\hat{\mathcal{A}}^{\mu_2...\mu_{p+1}}\partial_j\xi_A(y^1,y^2,...,y^n).
\end{eqnarray}

For the condition CCIII
to be satisfied, we must have
\begin{eqnarray}
    \xi_A(y^1,y^2,...,y^n) \sim \text{constant}.
\end{eqnarray}

With this, the $\mu_1-\nu_1$ component of $T_{MN}$ are
\begin{eqnarray}
    ^{(b)}T_{\mu_1\nu_1} = e^{-2p\sigma}\left[\frac{1}{p!}\hat{\mathcal{F}}_{\mu_1\mu_2...\mu_{p+1}}\hat{\mathcal{F}}_{\nu_1}^{\mu_2...\mu_{p+1}} - \frac{\hat{g}_{\mu_1\nu_1}}{2(p + 1)!}\hat{\mathcal{F}}_{\alpha_1...\alpha_{p+1}}\hat{\mathcal{F}}^{\alpha_1...\alpha_{p+1}}\right].
\end{eqnarray}

Furthermore, it is straightforward to check that
\begin{eqnarray}
  n\, ^{(b)}T_\alpha^\alpha - (d - 2)\, ^{(b)}T_j^j =  \frac{npe^{-2(p+1)\sigma}}{(p+1)!}\hat{\mathcal{F}}_{\mu_1\mu_2...\mu_{p+1}}\hat{\mathcal{F}}^{\mu_1\mu_2...\mu_{p+1}} 
\end{eqnarray}
and
\begin{eqnarray}
    ^{(b)}T_j^j = -\frac{n}{2(p + 1)!}e^{-2(p+1)\sigma}\hat{\mathcal{F}}_{\mu_1\mu_2...\mu_{p+1}}\hat{\mathcal{F}}^{\mu_1\mu_2...\mu_{p+1}} .
\end{eqnarray}
In which it becomes trivial to see that the only value of $p$ that satisfies all the consistency conditions presented here is $ p = 0$. This is in agreement with the conclusion of the authors of Refs. \cite{Duff:2000se,Freitas:2020vcf}. However, our result is stronger, as we show that only the 0-form is consistent, regardless of the number of dimensions and, more importantly, independent of the geometry of the internal space. Therefore, all models that consider the localization of $p$-forms with $p\neq 0$ must be revised, including models that aim to localize the 1-form and its dual in a string-like scenario, where the line 6D element it is a specific case of our general line element with $g_{\mu\nu} = e^{-2kr}\eta_{\mu\nu}$ and $g_{ik} = \text{diag}(e^{2B(r)},1)$ and is given by
\begin{eqnarray}
    ds^2 = e^{-2kr}\eta_{\mu\nu}dx^\mu dx^\nu + e^{-2B(r)}d\theta^2 + dr^2,
\end{eqnarray}
which is known to be a generalization of the models \cite{Gherghetta:2000jf,Cohen:1999ia}. This metric is used to localize fields in Refs.~\cite{Oda06,Oda03,Oda:2000kh,Alencar:2010vk}.
Our results show that all these mechanisms must be ruled out. In this sense, several other $p$-form field localization models in 6D and higher dimensions must be ruled out, such as in Refs.~\cite{Costa01,Costa02,Gogberashvili:2001jm}. The localization presented in \cite{Costa01} is performed considering that the line element is a specific case of our general element with $g_{\mu\nu} = e^{-kr + \tanh(kr)}\eta_{\mu\nu}$ and $g_{ik} = \text{diag}(e^{-kr + \tanh(kr)}\tanh^2(kr)k^{-2}, 1)$ and element line give by
\begin{eqnarray}
    ds^2 = e^{-kr + \tanh(kr)}(dx^\mu dx_\mu + \tanh^2(kr)k^{-2}d\theta^2) + dr^2,
\end{eqnarray}
which was presented in \cite{Silva:2012yj}.

\item \textbf{A no-go theorem for Nonlinear Electrodynamics (NED)}: The only localizable NED Lagrangian that satisfies all our consistency conditions is $L(F) = b\sqrt{F}$, where $b$ is a proportionality constant.

\textbf{Proof.}: Given the action of the NED field
\begin{equation}
    S^{(b)} = \int dx^d\, dy^n L(F),
\end{equation}
where we define, for convenience, $F$ as $F = F_{MN}F^{MN}$.

The the stress-energy tensor of the NED given by
\begin{eqnarray}
    ^{(b)}T_{MN} =  \frac{L}{2} g_{MN} - 2L_F F_{MQ}F_N^Q .
\end{eqnarray}

As previously done, we can write the vector $A_M = (A^T_\mu,0)$ and use separation of variables to express  $A^T_\mu = \hat{A}_\mu(x^\alpha)\xi_{NED}(y^1,y^2,...,y^n)$. With this, condition CCI implies that $\xi_{NED}(y^1,y^2,...,y^n)$ is constant. With this condition, we can write the $\mu-\nu$ component of the stress-energy tensor as
\begin{eqnarray}
    ^{(b)}T_{\mu\nu} =   \frac{L}{2} e^{2\sigma}\hat{g}_{\mu\nu} -2e^{-2\sigma}L_F \hat{F}_{\mu\alpha}\hat{F}_{\nu}^\alpha.
\end{eqnarray}

To satisfy the condition CCIII, we must therefore impose the following requirements
\begin{eqnarray}
\label{NED}    L &=& e^{-2\sigma}\hat{L}(x^\alpha),\\
    L_F &=& e^{2\sigma}\hat{L}_F(x^\alpha).
\end{eqnarray}

Finally, remembering that $F = e^{-4\sigma}\hat{F}$, the condition CCII implies that
\begin{eqnarray}
    L = 2FL_F = 2F\frac{dL}{dF},
\end{eqnarray}
whose solution is given by $L(F) = b\sqrt{F}$. Finally, Eq.~\eqref{NED} ensures that condition CCIV is automatically fulfilled, thereby completing and closing the proof.

\item \textbf{Vector field with scalar coupling}: An alternative would be to resort to the coupling models used in \cite{Kehagias,Fu:2011pu,Chumbes:2011zt,Landim03,Landim02,Landim01}, this time generalizing them to a braneworld of arbitrary dimension, where the usual gauge field Lagrangian is modified to
\begin{eqnarray}
    L(x,y) = -\frac{1}{4}f(y^1,y^2,...,y^n)F_{MN}F^{MN},
\end{eqnarray}
where the energy-momentum is now given by
\begin{eqnarray}
    T^{(b)}_{MN}(x,y) = f(y^1,y^2,...,y^n)\left(F_{MP}F_N^P - \frac{1}{4}g_{MN}F_{PQ}F^{PQ}\right).
\end{eqnarray}

Using the same gauge and variable separation as before, we have that the condition CCI implies again $\xi(y^1,y^2,...,y^n) \sim \text{constant}$. With this, we have that
\begin{eqnarray}
    ^{(b)}T_{\mu\nu} = e^{-2\sigma}f(y^1,y^2,...,y^n)\left(\hat{F}_{\mu\alpha}\hat{F}_\nu^{\alpha} - \frac{\hat{g}_{\mu\nu}}{4}\hat{F}_{\alpha\beta}\hat{F}^{\alpha\beta}\right),
\end{eqnarray}
them condition CCII implies that $f(y^1,y^2,...,y^n) = e^{2\sigma}$. However, if we apply condition CCIII, then we have
\begin{eqnarray}
    \frac{ne^{-4\sigma}f(y^1,y^2,...,y^n)}{2}\hat{F}_{\mu\nu}\hat{F}^{\mu\nu} = 0.
\end{eqnarray}

Therefore, even in braneworlds with arbitrary dimension and geometry, all these coupling models must be ruled out.

\item \textbf{Vector field with geometrical coupling}:

We may also consider the geometric couplings proposed in  \cite{Alencar02,Zhao,Alencar:2015rtc,Alencar03,Freitas}  in the context of braneworlds with arbitrary dimensions. Accordingly, by considering the energy-momentum tensor as
\begin{eqnarray}
        ^{(b)}T_{MN}&=&     \lambda_{1} \left[R_{MN} A_{P}A^{P} +R A_{M}A_{N}+ g_{MN} \Box(A_{Q}A^{Q}) - \nabla_{M}\nabla_{N}(A_{Q}A^{Q})\right]
        \nonumber \\
   && + \lambda_{2} \left[ 2R_{NQ}A_{M}A^{Q}-\nabla_{P}\nabla_{N}(A^{P}A_{M}) + \frac{1}{2} \Box(A_{M}A_{N}) +\frac{1}{2} g_{MN} \nabla_{P}\nabla_{Q}(A^{P} A^{Q}) \right]
   \nonumber \\
   && + F^{Q}_{M} F_{NQ} - g_{MN} \mathcal{L}_{matter},
\end{eqnarray}
we find that its $\mu-\nu$ component can be written as
\begin{eqnarray}
    ^{(b)}T_{\mu\nu} = e^{-2\sigma}\xi^2\hat{F}_{\mu\alpha}\hat{F}_{\nu}^{\alpha} +...
\end{eqnarray}

Therefore, in order to satisfy the condition CCIII, we must have $\xi = e^{\sigma}$. With this, the condition CCI reduces to
\begin{eqnarray}
    ^{(b)}T_{\mu j} = e^{-2\sigma}\xi \partial_j \xi B_{\alpha}\hat{F}_{\mu}^{\alpha} = 0.
\end{eqnarray}

Which is a contradiction. Therefore, all the models proposed in \cite{Alencar02,Zhao,Alencar:2015rtc,Alencar03,Freitas}  also fail in the context of arbitrary dimensions and must be ruled out.

\end{itemize}

\subsection{Fermionic fields}

\begin{itemize}
    \item \textbf{Dirac Fermions}: We will now work with the fermionic field in a codimension-one setup ($n=1$) with $d$ dimensions on the brane, which is described by the following Lagrangian:
    \begin{equation}
\mathcal{L}(x^{\nu},y) = \bar{\Psi}  i \Gamma^M D_M \Psi+ \lambda F(y) \bar{\Psi} \Psi\label{eq:lagrangiana},
\end{equation}
where $\Psi(x^{\nu},y)$ is our field and the covariant derivative is defined as $D_M = \partial_M + \omega_M$, with $\omega_M = \frac{1}{4} \omega_M^{AB} \Gamma_A \Gamma_B$, where $\omega_M^{AB}$ are the spin connections. In addition, we introduce a Yukawa-type interaction, where $\lambda$ is the coupling constant and $F(y)$ is a function that depends solely on the extra dimensions. The inclusion of this coupling term is necessary, in the literature, this mechanism is employed to confine one of the chiralities on the brane. We first analyze the case without the Yukawa term and then introduce it to verify that, even with its inclusion, the spinor fields do not satisfy the desired consistency conditions. We then proceed by considering the energy-momentum tensor
\begin{equation}
^{(b)}T_{MN}(x^\mu,y) = \frac{i}{2} \bar{\Psi} \Gamma_{(M} \nabla_{N)} \Psi - \frac{i}{2} \nabla_{(N} \bar{\Psi} \Gamma_{M)} \Psi  - \frac{1}{2} g_{MN} \lambda F(y) \bar{\Psi} \Psi .
\end{equation}

Unlike the bosonic case, the fermionic sector exhibits a crucial distinction when deriving the corresponding tensors. When dealing with spinors, we obtain one solution for the even-dimensional case $D=2k+2$ and another for the odd-dimensional case $D=2k+3$, where $D$ denotes the bulk dimension. This difference arises because spinors in $D$ dimensions transform under a representation of dimension $2^{[D/2]}$ \cite{Mendes}, which leads to significant modifications when performing dimensional reduction in order to obtain the effective spinors confined to the brane.

\end{itemize}

\subsubsection{Spinors in odd-dimensional spacetimes, $D=2k+3$}

For the odd-dimensional case, and in the absence of the Yukawa coupling, the spinor field in $D$ dimensions can be decomposed into an effective spinor living on the $d$-dimensional brane and a scalar profile that encodes the dependence of the field along the extra dimension \cite{Freitas:2022whn,Freitas:2020mxr,Alencar:2024lrl}.
\begin{equation}
    \Psi(x^{\nu},y)=\xi(y)\psi(x^\nu), \qquad\bar{\Psi}(x^{\nu},y)=\xi^{\dagger}(y)\bar{\psi}(x^\nu).
\end{equation}
With this type of decomposition, we obtain the following spin connections.
\begin{equation}
    \omega_\mu(x^{\nu},y)=\hat{\omega}_\mu(x^\nu)+\frac{1}{2}\sigma'(y)\Gamma_\mu(x^{\nu},y)\Gamma^y(x^{\nu},y), \qquad \omega_y(x^{\nu},y)=\hat{\omega}_y(y),
\end{equation}
and our gamma matrices are given by:
\begin{equation}
    \Gamma^\mu(x^\nu,y) = e^{-\sigma(y)}\,\hat{\Gamma}^\mu(x^\nu),
\qquad
\Gamma^y(x^\nu,y) = \Gamma^y(y).
\end{equation}

Consequently, the components of the energy-momentum tensor may be expressed as:
\begin{eqnarray}\label{relation_m1}
^{(b)}T_{\mu\nu} &=& \frac{i}{2}\,\xi^2 e^{\sigma}\Big[\bar{\psi}\,\hat{\Gamma}_{(\mu}\hat{\nabla}_{\nu)}\psi
-\hat{\nabla}_{(\nu}\bar{\psi}\,\hat{\Gamma}_{\mu)}\psi\Big],\\
\label{relation_m2}
^{(b)}T_{y\mu} &=& \frac{i}{4}\,\xi^2 
\Big[\bar{\psi}\,\hat{\Gamma}_{y}\hat{\nabla}_{\mu}\psi
-\hat{\nabla}_{\mu}\bar{\psi}\,\hat{\Gamma}_{y}\psi\Big],\\
\label{relation_m3}
^{(b)}T_{yy} &=& 0,\\
\label{relation_m4}
^{(b)}T_\mu{}^\mu-(d-2)\,^{(b)}T_y{}^y &=& \frac{i}{2}\,\xi^2 e^{-\sigma}\Big[\bar{\psi}\,\hat{\Gamma}^{\mu}\hat{\nabla}_{\mu}\psi
-\hat{\nabla}_{\mu}\bar{\psi}\,\hat{\Gamma}^{\mu}\psi\Big]\,.
\end{eqnarray}

From this analysis, we see that even for the zero mode of the fields, Eq.~\eqref{relation_m2} fails to satisfy the consistency condition~\eqref{condition_1} , except through the trivial solution \(\psi = \text{constant}\). Therefore, we conclude that the free spinor field is not localized.

\vspace{0.3cm}

By introducing the Yukawa coupling term 
\(\lambda F(y)\,\bar{\Psi}\Psi\) \cite{Alencar:2014moa}, thereby recovering our initially proposed Lagrangian 
\eqref{eq:lagrangiana}, we obtain the following contributions to the energy-momentum tensor:
\begin{equation}\label{eq:final}
\,^{(b)}T_{\mu\nu}=\frac{i}{2}\,\xi^2 e^{\sigma}\Big[\bar{\psi}\,\hat{\Gamma}_{(\mu}\hat{\nabla}_{\nu)}\psi-\hat{\nabla}_{(\nu}\bar{\psi}\,\hat{\Gamma}_{\mu)}\psi\Big]-\frac{1}{2}\xi^2 e^{2\sigma}\lambda \, \hat{g}_{\mu\nu}\,F(y)\,\bar{\psi}\psi\,,
\end{equation}
thus, in order to satisfy condition \eqref{geovamunu}, we take 
$\xi = e^{-\sigma/2}$ and $F(y) = C\,e^{-\sigma}$.
With this choice, we obtain
\begin{equation}\label{eq:final1}
\,^{(b)}T_{\mu\nu}=\frac{i}{2}\,\Big[\bar{\psi}\,\hat{\Gamma}_{(\mu}\hat{\nabla}_{\nu)}\psi-\hat{\nabla}_{(\nu}\bar{\psi}\,\hat{\Gamma}_{\mu)}\psi\Big]-\frac{1}{2}\,\lambda C \hat{g}_{\mu\nu}\,\bar{\psi}\psi\,.
\end{equation}
For the component \( ^{(b)}T_{yy} \), the relation \eqref{geova55} is satisfied
\begin{align}\label{eq:final2}
^{(b)}T_{yy} &= -\frac{1}{2} e^{-2A}\lambda C\bar{\psi}\psi\,.
\end{align}
For the trace, we have
\begin{align}\label{eq:final3}
\,^{(b)}T_\mu{}^\mu= \frac{i}{2}\,e^{-2\sigma}\Big[\bar{\psi}\,\hat{\Gamma}^{\mu}\hat{\nabla}_{\mu}\psi-\hat{\nabla}_{\mu}\bar{\psi}\,\hat{\Gamma}^{\mu}\psi\Big]-\frac{d}{2}e^{-2\sigma}\lambda C \,\bar{\psi}\psi, \
\end{align}
where, from the Dirac equation, we obtain the following relation
\begin{align}\label{eq:final4}
 \frac{i}{2}\,\Big[\bar{\psi}\,\hat{\Gamma}^{\mu}\hat{\nabla}_{\mu}\psi-\hat{\nabla}_{\mu}\bar{\psi}\,\hat{\Gamma}^{\mu}\psi\Big]=\lambda C \,\bar{\psi}\psi. \
\end{align}
Thus,
\begin{align}\label{eq:final5}
\,^{(b)}T_\mu{}^\mu= (1-\frac{d}{2})e^{-2A}\lambda C \,\bar{\psi}\psi,\
\end{align}
and 
\begin{align}\label{eq:final6}
\,^{(b)}T_y{}^y= -\frac{1}{2}e^{-2A}\lambda C \,\bar{\psi}\psi, \
\end{align}
and our relation \eqref{geovatraco} is satisfied.
\begin{align}\label{eq:final7}
\,^{(b)}T_\mu{}^\mu-(d-2)T_y{}^y &= 0.
\end{align}
For the component \( ^{(b)}T_{y\mu} \), since \( g_{y\mu} = 0 \), we obtain the same term as before, for which there is no way to satisfy relation \eqref{condition_1}
\begin{align}\label{eq:final8}
^{(b)}T_{y\mu} &= \frac{i}{4}\,\xi^2 
\Big[\bar{\psi}\,\hat{\Gamma}_{y}\hat{\nabla}_{\mu}\psi
-\hat{\nabla}_{\mu}\bar{\psi}\,\hat{\Gamma}_{y}\psi\Big].
\end{align}
From this analysis, we see that even when the fermionic field is coupled to a Yukawa term in a codimension-one model with \( d \) dimensions on the brane, the consistency relations proposed in \cite{Alencar:2024lrl} are not satisfied. Consequently, this localization mechanism cannot account for the confinement of fermionic fields.

\subsubsection{Spinors in even-dimensional spacetimes, $D=2k+2$}

In the even-dimensional case, we cannot use the same spinor decomposition adopted for odd values of $D$. When performing the dimensional reduction to $d = D-1$, the effective spinor must change its number of components, since the spinorial representation of $SO(1,D-1)$ is reducible (it admits a chiral decomposition), whereas the representation of $SO(1,d-1)$is irreducible. As a result, the reduction requires combining (or eliminating) the components associated with the two chiralities in order to obtain a spinor whose dimension matches the representation in $d$ dimensions. In this case, the spinor field in $D$ dimensions must take the following form :
\begin{equation}\label{eq:split}
\Psi(x^\nu,y) = \psi(x^\nu)\,\otimes\xi(y),
\qquad
\bar{\Psi}(x^\nu,y) = \bar{\psi}(x^\nu)\,\otimes({\xi^{\dagger}(y)}\sigma_1).
\end{equation}
and the corresponding spin connections are given by: 
\begin{equation}\label{eq:conn}
\omega_\mu(x^\nu,y)=\hat{\omega}_\mu(x^\nu)+\frac{1}{2}\,\sigma'\,\Gamma_\mu\,\Gamma^{y}\,,
\qquad
\omega_y(x^\nu,y)=0,
\end{equation}
where we have the following gamma matrices:
\begin{equation}\label{eq:gammas}
\Gamma^\mu(x^\nu,y)=e^{-\sigma}\,\hat{\Gamma}^\mu(x^\nu)\otimes\sigma_1,
\qquad
\Gamma^y(x^\nu,y)=\hat{\Gamma}^{y}(y)\otimes\sigma_3.
\end{equation}
where $\mu=0,....,D-2$, $y=D-1$, $\hat{\Gamma}^{y}(y)=\mathbf{I}$ and $\sigma_i$ denote the Pauli matrices. All these conditions hold for $D=2k+2$, where $D$ denotes the total bulk dimension. Consequently, the components of the energy-momentum tensor may be expressed as \cite{Mendes}:
\begin{eqnarray}
^{(b)}T_{\mu\nu}
&=& 
\frac{i}{2}e^{\sigma}\,\xi^\dagger\xi\Big[
 \bar{\psi}\,\hat{\Gamma}_{(\mu}\bar{\nabla}_{\nu)}\psi
 -\bar{\nabla}_{(\nu}\bar{\psi}\,\hat{\Gamma}_{\mu)}\psi
 \Big],
\\[0.2cm]
^{(b)}T_{y\mu}
&=& 
\frac{\xi^\dagger\sigma_{2}\,\xi}{4}\Big[
 \bar{\psi}\,\hat{\Gamma}_{y}\bar{\nabla}_{\mu}\psi
 -\bar{\nabla}_{\mu}\bar{\psi}\,\hat{\Gamma}_{y}\psi
 \Big],
\label{restriction_par}
\\[0.2cm]
^{(b)}T_{yy}
&=& 0,
\label{eq:Tyy}
\\[0.2cm]
^{(b)}T_{\mu}^{\mu}-(d-2)\,^{(b)}T_{y}^{y}
&=&
\frac{i}{2}\,\xi^\dagger\xi\, e^{-\sigma}
\Big[
\bar{\psi}\,\hat{\Gamma}^{\mu}\hat{\nabla}_{\mu}\psi
-\hat{\nabla}_{\mu}\bar{\psi}\,\hat{\Gamma}^{\mu}\psi
\Big].
\label{eq:final9}
\end{eqnarray}

Even with the change in approach to the spinor field in the even-\(D\) case, and adopting \(\xi^\dagger\xi = e^{-\sigma}\),
Eq.~\eqref{restriction_par} once again fails to satisfy relation~\eqref{condition_1}, 
as this would only occur in a trivial and highly constrained case, which is not of interest here. 
Consequently, it is again impossible to localize the free spinor field in a codimension-one setup 
with \(d\) dimensions on the brane for even \(D\).

By adding the Yukawa term, we obtain the following contribution to the energy--momentum tensor:
\begin{equation}
\Delta T_{MN} = -\frac{1}{2} g_{MN} \, \lambda \, F(y) \, (\xi^{\dagger} \sigma_{1} \xi) \, \bar{\psi} \psi .
\end{equation}
With this, we obtain the following equations:
\begin{eqnarray}
&& ^{(b)}T_{\mu\nu}
= 
\frac{i}{2}e^{\sigma}\xi^\dagger\xi\Big[
 \bar{\psi}\,\hat{\Gamma}_{(\mu}\bar{\nabla}_{\nu)}\psi
 -\bar{\nabla}_{(\nu}\bar{\psi}\,\hat{\Gamma}_{\mu)}\psi
 \Big]
 -\frac{1}{2} e^{2\sigma}\lambda\, \hat{g}_{\mu\nu} F(y)\,(\xi^{\dagger}\sigma_{1}\xi)\,\bar{\psi}\psi,
\\[0.25cm]
&& ^{(b)}T_{y\mu}
=
\frac{\xi^\dagger\sigma_{2}\,\xi}{4}\Big[
 \bar{\psi}\,\hat{\Gamma}_{y}\bar{\nabla}_{\mu}\psi
 -\bar{\nabla}_{\mu}\bar{\psi}\,\hat{\Gamma}_{y}\psi
 \Big],
\label{restriction_par2}
\\[0.25cm]
&& ^{(b)}T_{yy}
= 
-\frac{1}{2}\lambda\,F(y)\,(\xi^{\dagger}\sigma_{1}\xi)\,\bar{\psi}\psi,
\\[0.25cm]
&& ^{(b)}T_{\mu}^{\mu} - (d-2)\,^{(b)}T_{y}^{y}
=
\frac{i}{2}\,\xi^\dagger\xi\, e^{-A}
\Big[
\bar{\psi}\,\hat{\Gamma}^{\mu}\hat{\nabla}_{\mu}\psi
-\hat{\nabla}_{\mu}\bar{\psi}\,\hat{\Gamma}^{\mu}\psi
\Big]
-\lambda\,F(y)\,(\xi^{\dagger}\sigma_{1}\xi)\,\bar{\psi}\psi.
\label{eq:final9b}
\end{eqnarray}

 We see that even by choosing 
$\xi^\dagger\xi = e^{-\sigma},$
$\xi^\dagger \sigma_{1}\xi = e^{+\sigma},$  
and $F(y) = C e^{-\sigma}$, 
and by considering the zero mode of the fields, 
the Yukawa coupling is still not sufficient to satisfy relation~\eqref{condition_1}. 
This demonstrates that the even-\(D\) case also fails to satisfy the consistency relations, 
and that the localization mechanism does not work.

\section{Conclusion}\label{SecV}

In this work, we have presented a comprehensive and systematic analysis of field localization in braneworld scenarios with an arbitrary number of dimensions. Our approach builds upon and extends the framework introduced in Ref.~\cite{Alencar:2024lrl}, in which consistency conditions were derived to ensure that bulk fields are fully compatible with the Einstein equations in Randall-Sundrum-type geometries. A central advantage of this method is that it enables the investigation of localization properties directly at the level of the gravitational consistency conditions, without the need to invoke the specific equations of motion of the bulk fields. This makes the analysis both more general and broadly applicable to a wide class of braneworld models.

It was demonstrated which conditions the bulk energy-momentum tensors must satisfy. In particular, we showed that the previously known conditions, namely, that the $\mu-\nu$ component of the energy-momentum tensor should depend only on the brane coordinates and that the transverse component $\mu-j$ should vanish, remain unchanged in the context of braneworld models with arbitrary dimensions. On the other hand, we found that the conditions involving the trace of the energy-momentum tensor depend on the number of dimensions considered. In particular, we showed that any energy-momentum tensor must satisfy $n\,T_\alpha^\alpha = (d - 2)T_j^j$. This condition proved to be crucial in determining whether the fields studied here are consistent with the dynamics of the Einstein equation. Another interesting aspect worth mentioning is that none of these conditions depend, \textit{a priori}, on the geometry of the internal space, characterized by $g_{ik}$.

As an initial application of our formalism, we analyzed the energy--momentum tensor associated with a bulk scalar field. We showed that the local consistency conditions can only be satisfied when the scalar potential vanishes, $V(\Phi)=0$, implying that the scalar field must be free. Therefore, even within the fully general framework developed here, free scalar fields admit a consistent and normalizable zero mode, confirming their robustness with respect to dimensionality and background geometry.
We then established a no--go theorem for $p$-form fields, demonstrating that consistency and localization are simultaneously possible only for the $0$-form. This result is in agreement with earlier findings in Refs.~\cite{Duff:2000se,Freitas:2020vcf}. However, unlike those analyses, which were restricted to specific internal geometries, a limited number of dimensions, and relied explicitly on the equations of motion, our proof is entirely general. It applies to any smooth or regularized braneworld configuration described by a metric of the form~\eqref{metric}, independently of the dimensionality of spacetime or the structure of the internal space, whereas singular thin-brane configurations require a model-dependent prescription, what is not the objective of this work. As a direct consequence, we conclude that the localization mechanisms proposed in Refs.~\cite{Oda06,Oda03,Midodashvili,Oda:2000kh,Alencar:2010vk,Costa01,Costa02,Gogberashvili:2001jm,Oda05,Flachi,Choudhury:2006nj,Arkani-Hamed:1999wga} are not compatible with the full set of consistency conditions derived here.
In the same spirit, we showed that models aiming to localize vector fields through non-minimal couplings, either to scalar backgrounds or directly to geometric quantities, also fail to satisfy our local constraints. Consequently, the constructions proposed in Refs.~\cite{Kehagias,Fu:2011pu,Chumbes:2011zt,Landim03,Landim02,Landim01,Alencar02,Zhao,Alencar03,Freitas} must likewise be excluded, even when one allows for the most general braneworld setups.

Subsequently, we derived a general no-go theorem for nonlinear electrodynamics in Randall-Sundrum-type braneworld configurations. By applying the local consistency conditions to a completely arbitrary electromagnetic Lagrangian of the form $L(F)$, we found that nonlinear extensions of Maxwell theory generically fail to admit a normalizable and consistent zero mode on the brane. Remarkably, there is a single exception to this conclusion: the square-root model defined by $L(F)=b\sqrt{F}$. 
This result is fully general and does not rely on any particular choice of warp factor, dimensionality, or internal geometry. It follows purely from the local structure of the coupled gravitational and field equations. As a consequence, all localization mechanisms based on nonlinear deformations of Maxwell theory are excluded, unless the theory precisely reduces to the square-root form. In this sense, the model $L(F)=b\sqrt{F}$ emerges as the unique nonlinear electrodynamics that is both consistent with gravity and capable of supporting a localized zero mode in arbitrary--dimensional braneworld scenarios.

Turning to the fermionic sector, we find a scenario that closely parallels what occurs for scalar, gauge, and $p$-form fields. In particular, for fermionic fields propagating in codimension-one backgrounds with a $d$-dimensional brane, the requirements imposed by the no-go theorem are not fulfilled. This incompatibility persists even when one adopts one of the most widespread strategies in the literature, namely the inclusion of a Yukawa-type coupling between the fermion and background fields.
Despite its frequent use as a mechanism to localize at least a single chiral component of the fermionic zero mode, the Yukawa interaction fails to restore consistency once the full set of gravitational constraints is taken into account. As a result, the fermionic energy--momentum tensor cannot satisfy the local consistency relations required by the Einstein equations. We therefore conclude that this localization mechanism must be ruled out: fermionic fields, even with Yukawa couplings, cannot be consistently localized on the brane within the framework considered here.

Furthermore, the existence of a consistent localized NED solution suggests the possibility of finding black hole geometries within this framework. Indeed, we demonstrate that such solutions can be successfully constructed \cite{Alencar:2026dsq}. The feasibility of these solutions highlights that our framework inherently incorporates the gravitational backreaction. In the general line element of \eqref{metric}, the metric components and the warp factor are not fixed a priori, but are dynamic variables determined by the full $D$-dimensional Einstein equations. Within this structure, while the vacuum energy-momentum tensor governs the functional form of the warp factor and $\tilde{g}_{ij}(y)$, the bulk matter fields incorporate the backreaction by directly modifying and determining the brane metric. Consequently, different bulk tensors generate distinct brane geometries, a feature that arises from the specific decomposition of the Einstein tensor under a warped ansatz. In this sense, the local consistency conditions (CCs) derived in this work should be understood as integrability requirements. They define the physical constraints that bulk fields must satisfy to ensure that their backreaction remains compatible with the assumed metric symmetries of equation \eqref{metric}. Fields that violate these conditions, even if they result in a finite effective action, are not consistent with the Einstein equations through the symmetries of a warped geometry, necessitating a reconsideration of their localization mechanisms in such scenarios.

More generally, these results illustrate that the consistency conditions derived in this work sharply restrict the class of bulk fields that can support localized and well-defined compact objects in Randall-Sundrum-type scenarios. In particular, our framework singles out, in a unique and unambiguous way, which fields can be consistently confined to the brane while simultaneously guaranteeing that the resulting four-dimensional effective geometry is well defined and physically sensible.
We stress that these conclusions depend in an essential way on the \emph{local} nature of the consistency conditions imposed here, which encode the full interplay between bulk matter and gravitational dynamics at each point in the extra dimensions. From this perspective, it is natural to expect that the set of admissible localized fields could be enlarged in theories where the gravitational sector is modified. Extensions beyond standard Einstein gravity, such as higher-curvature corrections or scalar-tensor theories, may alter the local constraints and potentially allow new localization mechanisms. Exploring these possibilities lies beyond the scope of the present work and is left for future investigations.

\section*{Acknowledgments}
The authors would like to thank Conselho Nacional de Desenvolvimento Cient\'{i}fico e Tecnol\'ogico (CNPq), Coordena\c c\~ao de Aperfei\c{c}oamento de Pessoal de N\'{i}vel Superior (CAPES), and Funda\c c\~ao Cearense de Apoio ao Desenvolvimento Cient\'ifico e Tecnol\'ogico (FUNCAP) for the financial support.
FSNL acknowledges support from the Funda\c{c}\~{a}o para a Ci\^{e}ncia e a Tecnologia (FCT) Scientific Employment Stimulus contract with reference CEECINST/00032/2018, and funding through the research grants UID/04434/2025 and PTDC/FIS-AST/0054/2021.

\bibliographystyle{apsrev4-1}
\bibliography{ref}

\end{document}